\documentclass[a4paper,fleqn]{cas-sc}

\usepackage[numbers]{natbib}
\usepackage{cleveref}
\usepackage{mathtools}
\usepackage{subcaption}
\usepackage{svg}
\usepackage{amsmath}
\usepackage{soul}

\def\tsc#1{\csdef{#1}{\textsc{\lowercase{#1}}\xspace}}
\tsc{WGM}
\tsc{QE}
\tsc{EP}
\tsc{PMS}
\tsc{BEC}
\tsc{DE}

\newcommand{\rv}[1]{\textcolor{black}{#1}}

\begin{document}
\let\WriteBookmarks\relax
\def\floatpagepagefraction{1}
\def\textpagefraction{.001}

\shorttitle{On Micropolar Elastic Foundations}

\shortauthors{A. E. F. Athanasiadis et~al.}

\title [mode = title]{On Micropolar Elastic Foundations}                      



%
\author[1]{Adrianos E. F. Athanasiadis}






\affiliation[1]{organization={Institute for Infrastructure and Environment, The University of Edinburgh},
    addressline={William Rankine Building, Thomas Bayes Road, King's Buildings}, 
    postcode={EH9 3FG}, 
    city={Edinburgh},
    country={United Kingdom}}

\author[2]{Michal K. Budzik}

\author[1]{Dilum Fernando}

\affiliation[2]{organization={Department of Mechanical and Production Engineering, Aarhus University},
    addressline={Inge Lehmanns Gade 10, Navitas Building}, 
    postcode={8000}, 
    city={Aarhus C},
    country={Denmark}}

\author[1]{Marcelo A. Dias}[orcid=0000-0002-1668-0501]
\cormark[1]
\ead{marcelo.dias@ed.ac.uk}
\cortext[cor1]{Corresponding author}




\begin{abstract}
The modelling of heterogeneous and architected materials poses a significant challenge, demanding advanced homogenisation techniques. However, the complexity of this task can be considerably simplified through the application of micropolar elasticity. Conversely, elastic foundation theory is widely employed in fracture mechanics and the analysis of delamination propagation in composite materials. This study aims to amalgamate these two frameworks, enhancing the elastic foundation theory to accommodate materials exhibiting micropolar behaviour. Specifically, we present a novel theory of elastic foundation for micropolar materials, employing stress potentials formulation and a unique normalisation approach. Closed-form solutions are derived for stress and couple stress reactions inherent in such materials, along with the associated restoring stiffness. The validity of the proposed theory is established through verification using the double cantilever beam configuration. Concluding our study, we elucidate the benefits and limitations of the developed theory by quantifying the derived parameters for materials known to exhibit micropolar behaviour. This integration of micropolar elasticity into the elastic foundation theory not only enhances our understanding of material responses but also provides a versatile framework for the analysis of heterogeneous materials in various engineering applications.
\end{abstract}



\begin{keywords}
elastic foundation\sep micropolar elasticity \sep homogenisation \sep mechanical metamaterials\sep architected materials \sep heterogeneous materials
\end{keywords}

\maketitle

\section{Introduction}\label{sec1}

\rv{Heterogeneous and architected materials have garnered considerable attention due to their potential for tunability of mechanical properties and extensive use across many engineering applications~\citep{ashby2016}. Amongst these materials, it is encompassed a wide spectrum of functional structures, including low density cellular solids with extraordinary mechanical properties~\citep{zheng2014ultralight}, most of which requiring a clever combination of materials and geometry to a desirable behaviour. 
However, the complexity inherent in modelling these heterogeneous materials presents a substantial computational challenge. This primarily arises from complexity in geometric design and diverse material compositions in these structures~\citep{gibson1997,deshpande2001,shaikeea2022}.}


\rv{Given the increased interest in such architected systems, it seems suitable to seek for modelling approaches that ought to capture the complexity brought by heterogeneities in materials. Micropolar elasticity, as a theoretical framework, captures
microstructural kinematics, couple stresses, nonlocal behaviour, size dependence, and higher-order effects. Rooted in the pioneering work of the Cosserat brothers~\citep{cosserat1909} and subsequently revived by Eringen~\citep{eringen1968}, this theory introduces internal rotational degrees-of-freedom to each material point. These are known as microrotations and are distinct from that which is the classical infinitesimal rotations, thus allowing for modelling the changes in a set of material point directors. These directors represent the orientation of material elements at a specific point within a structure. This distinction between microrotations and infinitesimal rotations is a key feature of micropolar elasticity and has profound implications for capturing localised deformation behaviour. In essence, micropolar elasticity offers a unique and valuable perspective on materials by acknowledging the presence of internal structural elements within a material's constituents across the scales (e.g. grains in metals, cells in foams and fibres in fibre reinforced polymers). Furthermore, Nowacki's pioneering work~\citep{nowacki1986}, grounded in micropolar elasticity, has yielded valuable analytical results:
their research leveraged stress functions initially developed by Mindlin~\citep{mindlin1965}, enabling a systematic approach to solving problems of isotropic micropolar elasticity. 
In tandem with Nowacki's contributions, Lakes' studies~\citep{lakes1983,lakes1986,lakes1991} have further illuminated the practical applications of micropolar elasticity, particularly in elucidating the mechanical behaviour of foams and biological materials such as bones. A concrete framework for measuring micropolar material constants through pure tension, torsion and bending experiments has been developed by Gauthier~\citep{gauthier1975} and was utilised thoroughly through Lakes' works. Within this context, micropolar elasticity not only enhances our understanding of heterogeneous materials more broadly---\emph{e.g.}, geological materials are often considered as micropolar due to their layered structure~\citep{adhikary1997}---but also aids in the measurement of material constants, which are crucial for predicting and optimising their mechanical performance under different loading conditions.}

\rv{Significant advances have been made in utilising micropolar elasticity through the homogenisation of lattice structures, initially pioneered by Bazant~\citep{bazant1972}. This work notably reduced the computational cost associated with analysing large grid structures---a breakthrough later embraced by others~\citep{chen1998, wang1998, kumar2004, berkache2022} for modeling lattice structures and tackling fracture-related issues. Lattices conceptualised with beam elements have also been studied from an asymptotic homogenisation approach that led to effective micropolar continua~\citep{dosreis2012,alavi2022}. In a similar vain, dimensionally reduced beam theories, \emph{i.e.} Euler-Bernoulli and Timoshenko, have been augmented to include microrotational size effects~\citep{alavi2020}. Furthermore, on the manufacturing front, recent works showcase the importance of micropolar size effects in 3D printed lattices that present intense micropolar behaviour through high characteristic lengths for bending and torsion~\citep{ha2016,rueger2018}. These efforts and progress has broadened the practical applicability of micropolar elasticity, making it a powerful tool for understanding and predicting the behaviour of heterogeneous materials with complex internal structures.}

\rv{Parallel to these ideas, elastic foundation theory, originating from the pioneering work of Winkler~\cite{winkler}, introduces a fundamental concept that has revolutionised our approach to analysing confined structures and adhesive joints. At its core, elastic foundation theory replaces the continuum representation of a material with an array of non-interacting springs of infinitesimal size~\citep{dillard2018}. This conceptual shift allows for the treatment of complex structural configurations as interconnected springs, which collectively support the applied loads and govern deformation behaviour. The key advantage of this approach is its ability to address problems related to wrinkling and stable loading in confined structures. These issues are particularly relevant when dealing with thin faces bonded to bulk cores, as they frequently lead to interesting wrinkling phenomena~\cite{allen1969,audoly2008a}. An application of this is found in composites, which is used to optimise strength and mass distribution by combination of thin faces with bulk cores---for instance, better control of strength-to-weight ratios enhance fuel efficiency in aerospace applications~\citep{soutis2005}. Also crucial in renewable energy where composites contribute to lightweight, corrosion-resistant materials, thereby improving efficiency in turbine blades for wind and tidal infrastructure~\citep{thomas2018,alam2018}. However, thin faces, bulk cores and adhesive joints in composite structures, stress mismatches between adherends and the adhesive layer can lead to delamination and cracking, where differences in material properties result in uneven stress distribution, potentially compromising the joint's structural integrity; thus, optimising material compatibility and using appropriate adhesive thickness becomes crucial to mitigate these issues. Metamaterials, through their ability to provide tailored mechanical properties, engineered interfaces, and incorporate damage sensing functionalities, offer promising solutions for enhancing structural integrity and durability in composite materials~\citep{Athanasiadis2021,hedvard2024}.}

\rv{Elastic foundations have found extensive application in modelling cohesive fracture and delamination problems, particularly in the context of composite materials. For example, in the analysis of a Double Cantilever Beam (DCB) configuration, where two adherends are subjected to a bending load causing delamination or cohesive fracture, elastic foundation theory provides a simplified yet effective framework. This framework allows the interaction between lamellas (thin layers) to be represented as a beam resting on an elastic foundation. In doing so, it introduces a characteristic length-scale associated with a deformation zone, that is referred by the composites community as the fracture process zone, where delamination or cohesive fracture occurs~\citep{kanninen1973}. This approach has proven invaluable in characterising the mechanics of Mode I fracture, where crack growth occurs under tensile loading~\citep{heide2020}, Mode II fracture, where crack propagation takes place under shearing forces~\citep{budzik2013}, and mixed mode~\citep{bensalem2014}.} 

Despite these significant advancements in the fields of micropolar elasticity and elastic foundation theory, a notable research gap persists---a comprehensive framework that seamlessly integrates micropolar effects into elastic foundation theory is currently lacking. Furthermore, there is a shortage of research concerning the behaviour of micropolar elastic media under wrinkling-induced displacement loading. These gaps in knowledge limit our ability to fully leverage the potential benefits of micropolar elasticity in analysing complex confined structures. This article seeks to address these gaps by presenting a comprehensive theoretical framework for assessing the mechanical behaviour of micropolar elastic materials under diverse configurations. 

Our approach introduces a normalisation of the confinement problem, shedding light on the interplay of length scales that underlie the assumptions and providing a solid foundation for further exploration. Through this work we provide closed form solutions describing the deformation of micropolar elastic cores on the plane that are confined between thin elastic beams for various configurations of boundary conditions. Thereafter, we present an application of the developed theory in the DCB configuration, by deriving closed form solutions of the deflection curves and validating our findings against numerical solutions. Lastly, we demonstrate the limits of applicability of our theory by calculating the derived design parameters for the case of regular honeycomb core material, as well as Poly-Styrene (PS) and Poly-Urethane (PU) foams. 

In section 2 of this work, we present the fundamental definitions of planar micropolar elasticity of an isotropic medium in a dimensionless form and we layout the basic theory to be used. In section 3, we develop a theory for micropolar elastic foundations for examples of load configurations that can occur, where we assess the normal and rotational stiffness behaviour of each case studied. In section 4, we apply the developed theory on the DCB configuration whereas the developed theory is validated against numerical simulations.~\rv{Thereby, we observe a significant deviation on the DCB process zone morphology between our proposed model and classical Cauchy formalism (\emph{i.e.}, in the absence of micropolar contributions).} Lastly, we assess the significance and the limits of applicability of our method through scaling laws for selected material configurations with experimental data supplied from the available literature. 

\section{The Micropolar Continuum}\label{sec2}

\subsection{Definitions}\label{sec2.1}

\begin{figure}[t!]
    \centering
    \includegraphics[width=0.7\textwidth]{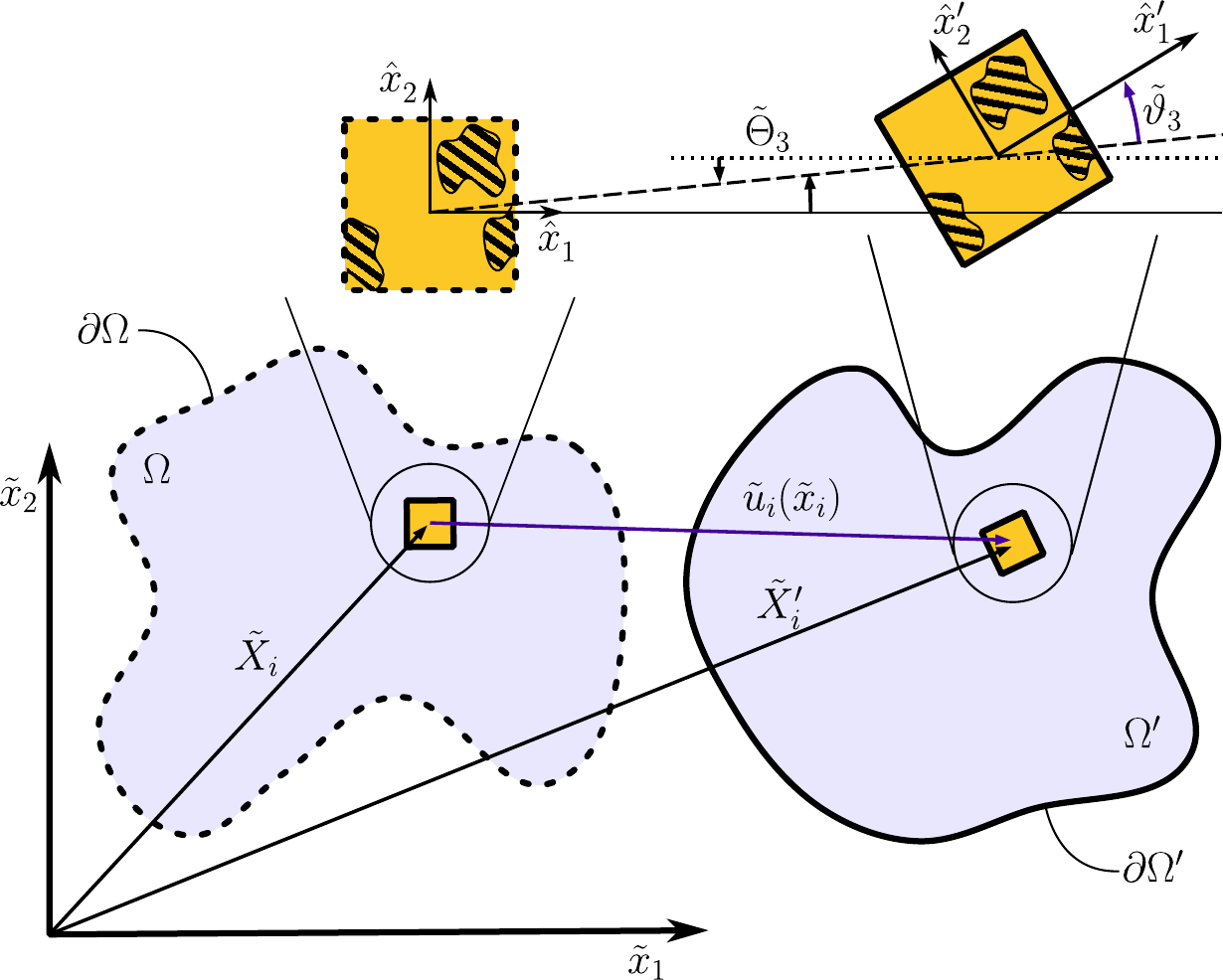}
    \caption{Kinematics of 2-Dimensional micropolar medium. An oriented material point in the reference configuration: $x_\alpha(x_\alpha)$ undergoes a displacement $u_\alpha(x_\alpha)$ resulting to infinitesimal rotation $\tilde{\Theta}_3=\epsilon_{\alpha\beta}\tilde{\partial}_\alpha\tilde{u}_\beta/2$. At the deformed configuration the new position of the material point is $X'_i=x_\alpha+u_i$. However, the local director of the material point may also experience a microrotation $\tilde\vartheta_3(\tilde{x}_\alpha)\neq\tilde\Theta_3$ that is distinct to the infinitesimal rotation.}
    \label{fig:definitions}
\end{figure}

We begin by setting up the theory of micropolar elasticity, which is here, for simplicity and justified by the study case presented in this article, aimed at 2--dimensions media. This is established by assuming that individual material points of an elastic body, both in the reference and the deformed configuration, have rotational degrees-of-freedom in addition to their displacements---these are known as microrotations, which are distinct to the infinitesimal rotations in standard elasticity~\citep{eringen1968}. We refer to~\cref{fig:definitions} to illustrate the kinematical variables. Therefore, let an arbitrary material point be tracked by Lagrangian coordinates denoted by $\tilde{x}_\alpha$\footnote{Noticed that variables with a tilde carry dimensions, which is length in the case of coordinates and displacements.}. The infinitesimal displacement fields associated to material points will be written as $\tilde{u}_\alpha(\tilde{x}_\alpha)$. However, in order to capture the internal degrees-of-freedom, directors associated to any material point may rotate, thus giving rise to a microrotation field denoted as $\tilde\vartheta_3(\tilde{x}_\alpha)$. This field is distinct from infinitesimal rotations, $\tilde\Theta_3(\tilde{x}_\alpha)=\epsilon_{\alpha\beta}\tilde\partial_\alpha\tilde{u}_\beta/2$, which are derived from the displacements ($\epsilon_{\alpha\beta}$ stands for the permutation symbol of $2^\text{nd}$ rank and $\tilde\partial_\alpha\equiv\partial/\partial\tilde{x}_\alpha$ stands for the spatial derivative acting at the $\alpha$ direction). The fundamental difference from a micropolar continuum to a classical one is that the contribution of microrotation to the strain tensor causes it to display a non-symmetric shear component. Hence, the non-symmetric strain field components is defined as follows:
\begin{equation}
\tilde\varepsilon_{\alpha\beta}=\frac{1}{2}\left(\tilde\partial_\alpha\tilde{u}_{\beta}+\tilde\partial_\beta\tilde{u}_{\alpha}\right)+\epsilon_{\alpha\beta}\left(\tilde\Theta_3-\tilde\vartheta_3\right)=\tilde\partial_\alpha\tilde{u}_{\beta}-\epsilon_{\alpha\beta}\tilde\vartheta_3.
\end{equation}
The existence of microrotations and their gradients (micro-curvatures) are balanced by couple stresses (micro-moments) that develop in the medium. Hence, the constitutive relations for an isotropic micropolar elastic material in 2--dimensions reads as~\citep{lakes1983,lakes1986}:
\begin{subequations}
     \label{eq:constitutive}
    \begin{align}
    \tilde\sigma_{\alpha\beta}&=(\mu+\kappa)\tilde\varepsilon_{\alpha\beta}+\mu\tilde\varepsilon_{\beta\alpha}+\lambda \tilde\varepsilon_{\gamma\gamma}\delta_{\alpha\beta},\\
    \tilde{m}_{\alpha 3}&=\gamma\tilde\partial_\alpha \tilde\vartheta_3,
    \end{align}
\end{subequations}
where, $\tilde\sigma_{\alpha\beta}$ and $\tilde{m}_{\alpha3}$ are the stress tensor and couple stress vector respectively. The symbols $\lambda$ and $\mu$ are the classical Lam\`e parameters, while $\kappa$ and $\gamma$ are related to the micropolar behaviour of the material~\citep{eringen1968,lakes1991}. The engineering moduli are recovered by the following relationships to the above constants:
\begin{subequations}
    \begin{align}
    E&=\frac{(2\mu+\kappa)(3\lambda+2\mu+\kappa)}{2\lambda+2\mu+\kappa},\\
    \nu&=\frac{\lambda}{2\lambda+2\mu+\kappa},\\
    \ell_b^2&=\frac{\gamma}{2(2\mu+\kappa)},\\
    N^2&=\frac{\kappa}{\kappa+(2\mu+\kappa)},
    \end{align}
\end{subequations}
where, $E$ stands for the Young's modulus, and $\nu$ is the Poisson's ratio. The new constant $\ell_b$ is a characteristic length for bending that measures the material's resistance to microcurvature and the coupling ratio $N$, which characterises the amount of coupling between linear and rotational degrees of freedom, is bounded by the inequality $0\leq N\leq1$. When $N=0$, the rotational stresses are uncoupled from the linear stresses, hence, classical elasticity is recovered. We also define the shear modulus as $2G=2\mu+\kappa=E/(1+\nu)$, which will be later used as convenient normalisation parameter.~\rv{Notice that, in this context, the Lam\`e parameter $\mu$ does not coincide with the shear modulus $G$, with the latter being the modulus one would measure during a standard torsion test~\citep{lakes1991}.}

To achieve static equilibrium, the balance equations for a micropolar body in 2--dimensions with absence of body distributed loads are: 
\begin{subequations}
    \begin{align}    
    \tilde\partial_\beta\tilde\sigma_{\beta\alpha}&=0\\
    \tilde\partial_\alpha\tilde{m}_{\alpha3}+\epsilon_{\alpha\beta}\tilde\sigma_{\alpha\beta}&=0.
    \end{align}
\end{subequations}
The above balance equations can be automatically satisfied by an appropriate choice of stress functions $\tilde\Phi(\tilde{x}_\alpha)$ and $\tilde\Psi(\tilde{x}_\alpha)$~\citep{nowacki1986,mindlin1965}, which must hold the following relationship to the stress and couple-stress fields: 
\begin{subequations}
 \label{eq:stressdefinitions}
    \begin{align}
\tilde\sigma_{\alpha\beta}&=\epsilon_{\alpha\gamma}\epsilon_{\beta\delta}\tilde\partial_\gamma\tilde\partial_\delta \tilde\Phi-\epsilon_{\alpha\gamma}\tilde\partial_\beta\tilde\partial_\gamma \tilde\Psi\\
    \tilde{m}_{\alpha3}&=\tilde\partial_\alpha \tilde\Psi.
    \end{align}
\end{subequations}
Notice that when $\tilde{\Psi}=0$ the usual Airy's stress function formulation is recovered. The compatibility conditions can be written independently of the material's constitutive behaviour when written on the strain fields:
\begin{equation}
            \epsilon_{\beta\gamma}\tilde{\partial}_\gamma\tilde\varepsilon_{\beta\alpha}+\tilde\partial_\alpha\tilde{\vartheta}_3=0.\label{eq:compatibility0}
\end{equation}
Taking a derivative and rearranging, \cref{eq:compatibility0} can be rewritten in the form:
\begin{subequations}
    	\begin{align}
    		\epsilon_{\alpha\gamma}\epsilon_{\beta\delta}\tilde\partial_\alpha\tilde\partial_\beta \tilde\varepsilon_{\gamma\delta}&=0\\
        		\epsilon_{\alpha\gamma}\tilde\partial_\beta\tilde\partial_\gamma\tilde\varepsilon_{\alpha\beta}+\tilde\partial_\alpha\tilde\partial_\alpha\tilde\vartheta_3&=0.
    	\end{align}
    \label{eq:compatibility}
\end{subequations}
However, since the stress potentials are defined through the stress fields, a constitutive law must be chosen to express compatibility through stress potentials. Upon substitution of \cref{eq:stressdefinitions} into the isotropic constitutive behaviour of~\cref{eq:constitutive}, and relating the resulting expression to the compatibility conditions in \cref{eq:compatibility}, we arrive at the differential equations that govern an isotropic micropolar elastic 2--dimensional solid, which simplify to: 
\begin{subequations}
    \begin{align}
    \tilde{\Delta}^2\tilde\Phi&=0\\
    \tilde{\Delta}(1- \omega^2\tilde{\Delta})  \tilde{\Psi}&=0,\label{eq:PDE}
    \end{align}
    \label{eq:biharmonic}
\end{subequations}
where,
\begin{equation}
    \omega=\frac{\ell_b}{N}=\sqrt{\frac{\gamma}{\kappa}\frac{\mu+\kappa}{2\mu+\kappa}}
\end{equation}
is a characteristic length scale associated to the material properties. The step from~\cref{eq:compatibility0} to~\cref{eq:compatibility} increases the order of the system, resulting to an additional integration constant to be determined. Therefore,~\cref{eq:compatibility0} will need to be subsequently satisfied. Furthermore,~\cref{eq:compatibility0} in terms of stress potentials for isotropic micropolar media takes the form of Cauchy-Riemann conditions~\citep{nowacki1986}, and reads as: 
\begin{equation}
    (1-\nu)\ell_b^2\tilde\partial_\alpha\tilde\Delta\tilde\Phi+\epsilon_{\beta\alpha}\tilde\partial_\beta(1-\omega^2\tilde\Delta)\tilde\Psi=0
\end{equation}
Given that the choice of constitutive behaviour is of an isotropic material, as shown in~\cref{eq:constitutive}, the resulting Partial Differential Equation (PDE) system for the stress potentials becomes uncoupled as in~\cref{eq:biharmonic}. On the other hand, if the material is no longer isotropic, the PDE system derived from the compatibility equations, \cref{eq:compatibility}, will be a linear $4^{\text{th}}$ order coupled system. It is observed that if we were to consider a micropolar behaviour that originates from planar lattice structures, we would need a unit-cell preserving 6-fold symmetries in order to have an isotropic medium~\citep{christensen1987}.

\subsection{Micropolar Elasticity in dimensionless form}

We proceed with a normalisation of the micropolar elasticity equations. Let $\ell$ be a characteristic length-scale associated to the typical size of the deformation decay of an applied external load or displacement. This length-scale normalises variables with dimensions of length as follows: $x_\alpha\equiv\tilde{x}_\alpha/\ell$ and $u_\alpha\equiv \tilde{u}_\alpha/\ell$. The normalised $n$-th order derivatives are hence written as $\partial^n_\alpha\equiv\ell^n\tilde\partial_\alpha^n=\ell^n\partial^n/\partial \tilde{x}_\alpha^n$. The double of the shear modulus is chosen as a reference force per unit of length square value, hence normalising stress fields as: $\sigma_{\alpha\beta}\equiv\tilde\sigma_{\alpha\beta}/(2G)$ and $m_{\alpha}\equiv\tilde{m}_{\alpha3}/(2G\ell)$. As a consequence of \cref{eq:stressdefinitions} and the chosen normalisation, we may write the dimensionless stress potentials as follows: $\Phi\equiv\tilde\Phi/(2G\ell^2)$ and $\Psi\equiv\tilde{\Psi}/(2G\ell^2)$. Furthermore, strain measures and rotations remain dimensionless, which are from this point forward referred to as $\varepsilon_{\alpha\beta}\equiv\tilde\varepsilon_{\alpha\beta}$ and $\vartheta\equiv\tilde\vartheta_3$. Based on these definitions, the normalised constitutive relations for plane strain and plane stress are respectively written using Voigt notation as~\citep{nakamura1995}: 
\begin{equation}
    \begin{pmatrix}
    \sigma_{11}\\ \sigma_{22}\\ \sigma_{12}\\ \sigma_{21} \\m_1 \\m_2
    \end{pmatrix}=\left[ \begin{array}{cccccc}
    \nu^* & \nu^*-1 & 0 & 0 & 0 & 0 \\
    \nu^*-1 & \nu^* & 0 & 0 & 0 & 0  \\
    0 & 0 & \frac{\nu^*/2}{1-N^2} & \frac{(\nu^*/2)(1-2N^2)}{1-N^2} & 0 & 0 \\ 
    0 & 0 & \frac{(\nu^*/2)(1-2N^2)}{1-N^2}& \frac{\nu^*/2}{1-N^2} & 0 & 0 \\
    0 & 0  & 0 & 0 & g & 0\\
    0 & 0  & 0 & 0 & 0 & g
    \end{array}\right]
    \begin{pmatrix}
    \varepsilon_{11}\\ \varepsilon_{22}\\ \varepsilon_{12}\\ \varepsilon_{21}\\ \partial_1\vartheta\\ \partial_2\vartheta
    \end{pmatrix},\label{eq:constitutiveplstrain}
\end{equation}
where 
\begin{equation}
    \nu^*\equiv
    \begin{dcases}
    (1-\nu)/(1-2\nu)& \text{if plane strain}\\
    1/(1-\nu)& \text{if plane stress}
    \end{dcases},
\end{equation}
and, if we define the dimensionless parameter $\eta^2\equiv \ell^2/\omega^2+1$, the micropolar parameter $\gamma$ is normalised as:
\begin{equation}
    g\equiv\frac{\gamma}{2G\ell^2}=\frac{2N^2}{\eta^2-1}=\frac{2\ell^2_b}{\ell^2}.
\end{equation}

\section{Elastic foundation formulation}

\begin{figure}
    \centering
    \includegraphics[width=0.7\textwidth]{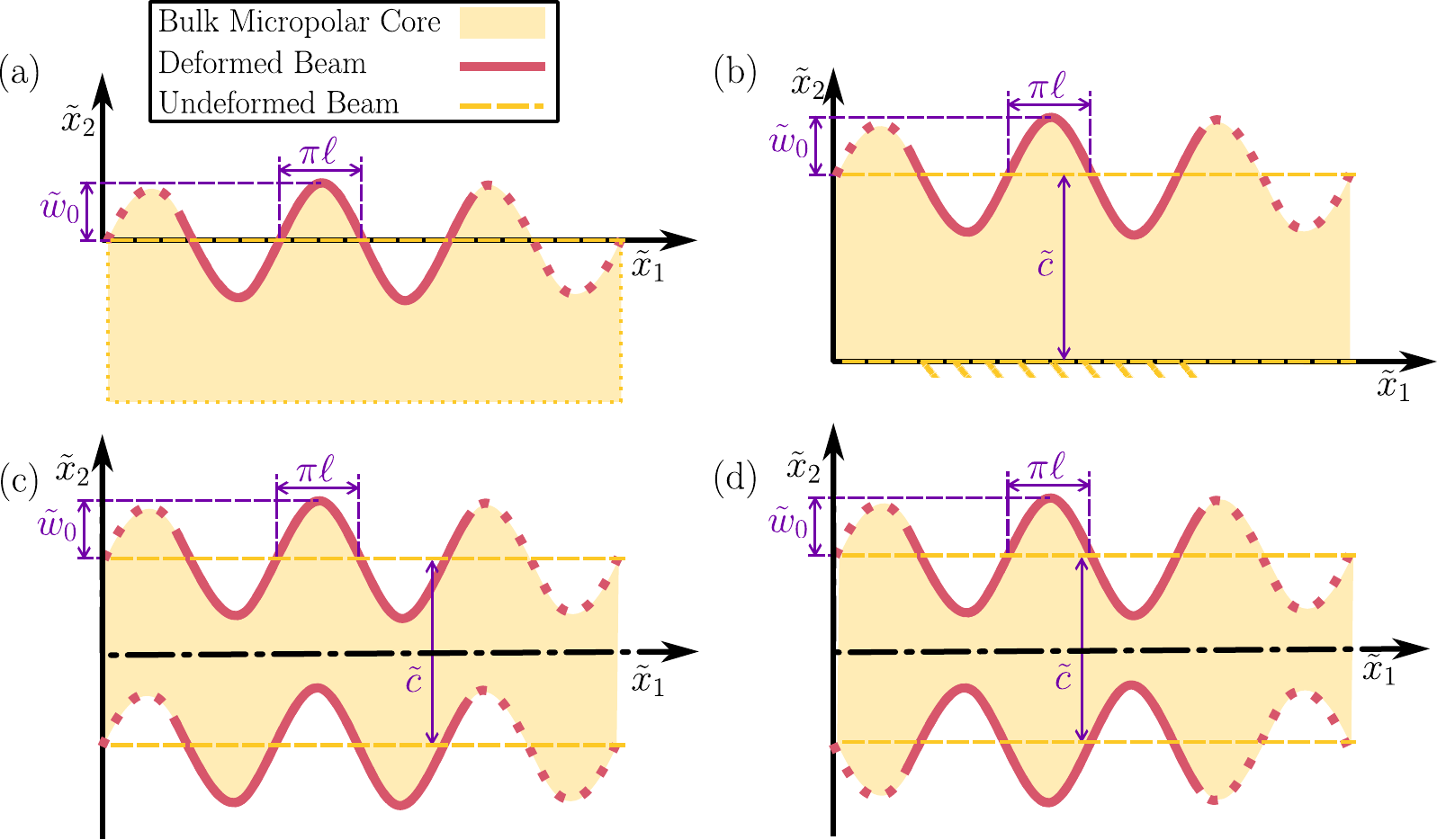}
    \caption{Schematic of the studied cases: micropolar core is attached to sinusoidally deformed beams. The amplitude and half-wavelength of the deformation profiles are $w_0$ and $\pi\ell$ respectively. That results to sinusoidally deformed top and/or bottom boundaries along the horizontal direction of the core. Illustration of the studied cases: (a) Infinitely thick foundation laying at the lower Euclidean semi-infinite plane. The interface between bulk and beam is located along the line $\tilde{x}_2=0$. (b) Foundation of finite thickness that is fixed at $\tilde{x}_2=0$ while the deformed boundary is located at $\tilde{x}_2=\tilde{c}$. (c) Foundation of finite thickness that is bonded between two boundaries that deform anti-symmetrically, located at $\tilde{x}_2=\pm \tilde{c}/2$. (d) Foundation of finite thickness that is bonded between two boundaries that deform symmetrically, located at $\tilde{x}_2=\pm \tilde{c}/2$. In all cases, the core extends infinitely in the $\pm\tilde{x}_1$ directions.}
    \label{fig:cases}
\end{figure}

In this section, we formulate a theory for micropolar elastic foundations to analyse load configurations that may arise, specifically addressing cases where the foundation presents some degree of complexity in their microstructure. We shall evaluate the normal, shear, and rotational stiffness characteristics of a foundation's behaviour, demonstrating their relevance and connection to microscopic architectures. Additionally, we will show the applicability and limitations of these stiffness characteristics in more general scenarios where micropolar elasticity applies, providing a comprehensive understanding of the foundation's mechanical response. Similarly to the existing theory for the classical elastic foundation~\cite{allen1969,hoff1945,gough1940}, we use a stress function formulation to describe the mechanical behaviour of the micropolar elastic core under four distinct sets of boundary conditions. Namely, semi-infinite foundation, a first case of confinement between a rigid and a deforming boundary, and two other cases of confinement by symmetrically and anti-symmetrically deforming boundaries, as shown in~\cref{fig:cases}. We assume that the boundary is deformed into a sinusoidal shape, $\tilde{w}_b(\tilde{x}_1)=\tilde{w}_0\sin(\tilde{x}_1/\ell)$, where $w_0$ and $\pi\ell$ are the amplitude and half-wavelength of the deformation profile, respectively. In dimensionless form, a boundary's deflection is written as: $w_b(x_1)=w_0\sin(x_1)$, where $w_0\equiv \tilde{w}_0/\ell$. We also define the dimensionless amplitude: $w_c=w_0/c=\tilde{w}_0/\tilde{c}$ for cores of finite thickness, where $c\equiv \tilde{c}/\ell$ is the dimensionless core thickness and $\tilde{c}$ is its dimensional counterpart. 

We now seek to measure the resulting restoring stress that is applied from the core to the beam. To achieve that, we must solve the bi-harmonic system given by \cref{eq:biharmonic}. This is done via separation of variables of the stress functions, which in their dimensionless versions take they following forms:
\begin{subequations}
    \begin{align}
    \Phi(x_\alpha)&=\phi(x_2)\sin\left(x_1\right)\\
    \Psi(x_\alpha)&=\psi(x_2)\cos\left(x_1\right).
    \end{align}  
\end{subequations}
Such consideration ensures that the boundaries will deform into the required sinusoidal profiles. Here, an essential assumption is established, wherein the foundation boundary surface functions as an interface between a micropolar medium and a different classical elastic medium (such as a beam resting on a micropolar foundation). Therefore, we must assume that at the interface there is no distinction between micrortations and infinitesimal rotations. Hence, the microrotation at the interfaces $\partial\Omega_\textrm{interface}\subseteq\partial\Omega$ will be equal to the infinitesimal rotation of the medium: $\vartheta=\epsilon_{\alpha\beta}\partial_\alpha u_\beta/2$ at $x_\alpha\in\partial\Omega_\textrm{interface}$, which justifies the selection for the $\Psi$ stress function. Upon substitution in \cref{eq:biharmonic}, the differential equations that describe the foundation are:
\begin{subequations}
\label{eq:ode_system}
    \begin{align}
    \label{eq:ode_system1}
    \phi^{(4)}(x_2)-2  \phi''(x_2)+\phi(x_2)&=0\\
    \label{eq:ode_system2}
   \psi ^{(4)}(x_2)-\left(1+\eta^2\right) \psi''(x_2)+\eta^2\psi (x_2)&=0.
    \end{align}
\end{subequations}
The above \cref{eq:ode_system} results in an uncoupled ordinary differential equation system of the fourth order along the thickness of the isotropic micropolar foundation. The general solutions of~\cref{eq:ode_system1,eq:ode_system2} read as follows:
\begin{subequations}
\begin{align}
    \phi(x_2)&=\left(c_1+c_2 x_2\right)\cosh{x_2}+\left(c_3+ c_4 x_2\right)\sinh{x_2}\\
    \psi (x_2)&=c_5\cosh{x_2}+c_6\cosh{\eta x_2}+c_7\sinh{x_2}+c_8\sinh{\eta x_2}.\label{eq:sols}
\end{align}
\end{subequations}
We relate the constants of integration via application of Cauchy-Riemann conditions, which results in the following relationship: 
\begin{equation}
    \frac{c_2}{c_5}=\frac{c_4}{c_7}=\frac{1}{2g}\frac{2\nu^*-1}{\nu^*}.\label{eq:cauchy}
\end{equation}
We consider four representative loading cases. Namely, infinitely thick foundation, finite foundation with rigid base, finite foundation with antisymmetric boundaries and finite foundation with symmetric boundaries. All the cases are schematically shown in~\cref{fig:cases}. The foundation always extends infinitely through the $\tilde{x}_1$ axis while the applied deformation profile is always chosen to be of sinusoidal shape. For all the cases studied, we assume that at the interface between micropolar foundation and the applied sinesoidal deformation, the foundation is locally classical elastic. Hence, one necessary boundary condition will be the matching of microrotation and infinitesimal rotation at the interface: $2\vartheta=\epsilon_{\alpha\beta}\partial_\alpha u_\beta$. This fact eliminates two of the constants for all cases studied: $c_6=c_8=0$. Therefore, the stress functions that describe the general problem, and after application of~\cref{eq:cauchy}, reduce as: 
\begin{subequations}
    \begin{align}
    \Phi(x_\alpha)&=\left(c_1 -c_5\frac{1-2\nu^*}{2g\nu^*} x_2\right)\sin x_1 \cosh x_2 +\left(c_3 - c_7\frac{1-2\nu^*}{2g\nu^*} x_2\right) \sin x_1 \sinh x_2\\
    \Psi(x_\alpha)&=(c_5  \cosh x_2 + c_7 \sinh x_2 )\cos x_1
    \end{align}
 \end{subequations}
Upon substitution at the stress definitions in~\cref{eq:stressdefinitions}, and the constitutive equations in~\cref{eq:constitutiveplstrain}. The displacement fields can be obtained by integration of the normal strains. That results to the following general displacement expressions:
 \begin{subequations}
    \begin{align}
    u_1(x_\alpha)&=\frac{\cos x_1}{2g\nu^*}\left\{\cosh x_2\left[c_5 \left(1-2 \nu ^*\right)-2 \nu ^* \left(g c_1+g c_7 +c_7\right) x_2\right]+\right.\nonumber\\
    &\quad\quad\left.+\sinh x_2\left[c_7 \left(1-2 \nu ^*\right) x_2-2 \nu ^* \left(g c_3+g c_5+c_5\right)\right]\right\}\\
        u_2(x_\alpha)&=\frac{\sin x_1}{2g\nu^*}\left\{\cosh x_2\left[c_7 x_2+c_5-2 \nu ^* \left(g c_3+g c_5+c_7 x_2\right)\right]+\right.\nonumber\\
    &\quad\quad\left.+\sinh x_2\left[c_5 x_2+c_7-2 \nu ^* \left(g c_1+g c_7+c_5 x_2\right)\right]\right\}\\
    \vartheta(x_\alpha)&=\frac{1}{g}\cos x_1\left(c_7 \sinh x_2+c_5 \cosh x_2\right).
    \end{align}\label{eq:displacements}
 \end{subequations}
We then calculate the reaction linear and couple tractions for the respective linear and rotational springs at the $x_2$ direction with respect to the dimensionless amplitude $w_0$. However, for cases with finite foundation thickness it is convenient to calculate stiffness with respect to the dimensionless amplitude $w_c$, hence we write $w_0=c w_c$ and express the stiffness as $k_i'=c k_i$---in this way the amplitude is independent of the parameter $c$.

The stress and strain fields are recovered by differentiating as per the definitions in~\cref{eq:stressdefinitions} and by using the constitutive relations of~\cref{eq:constitutiveplstrain} respectively. These are here expressed as follows: 
\begin{subequations}
    \begin{align}
    \sigma_{22}(x_1)&=k_n w_0\sin(x_1)=k_n' w_c\sin(x_1),\\
    \sigma_{21}(x_1)&=k_s w_0\cos(x_1)=k_s' w_c\cos(x_1),\\
    m_{23}(x_1)&=k_\vartheta w_0\cos(x_1)=k_\vartheta' w_c\cos(x_1).    
    \end{align}\label{eq:stiffness_def}
\end{subequations}
The deflection fields and the microrotation field of~\cref{eq:displacements} will contain the amplitude $w_0$ through the integration constants after application of the boundary conditions. Also, every displacement field as well as the microrotation field will be proportional to a trigonometric function. Therefore, we decide to express the solutions of~\cref{eq:displacements} through functions of the across-the-thickness coordinate $x_2$. We call those amplitude functions and we define them as:
 \begin{subequations}
    \begin{align}
    u_1(x_\alpha)&=w_0\hat{u}_1(x_2)\cos x_1,\\
    u_2(x_\alpha)&=w_0\hat{u}_2(x_2)\sin x_1,\\
    \vartheta(x_\alpha)&=w_0\hat{\vartheta}(x_2)\cos x_1.    
    \end{align}
 \end{subequations}
We then present the shapes of the amplitude functions: $\hat{u}_\alpha(x_2)$ and $\hat{\vartheta}(x_2)$, demonstrating their respective trends through the parametric domain. A representative selection for the values of the normalised core thickness is chosen as: $c=\{10^{1/2},10,10^{3/2}\}$ while the amplitude functions are drawn for the limits values of Poisson's ratio for both plane stress and plane strain conditions. Furthermore, we present the asymptotic behaviours for $c\rightarrow0$ and $c\rightarrow\infty$. All our results are validated through numerical results through finite element modelling of the respective cases. 
 
\subsection{Case A: Infinite core foundation}
\begin{figure}[htb!]
    \centering
    \includegraphics[width=\textwidth]{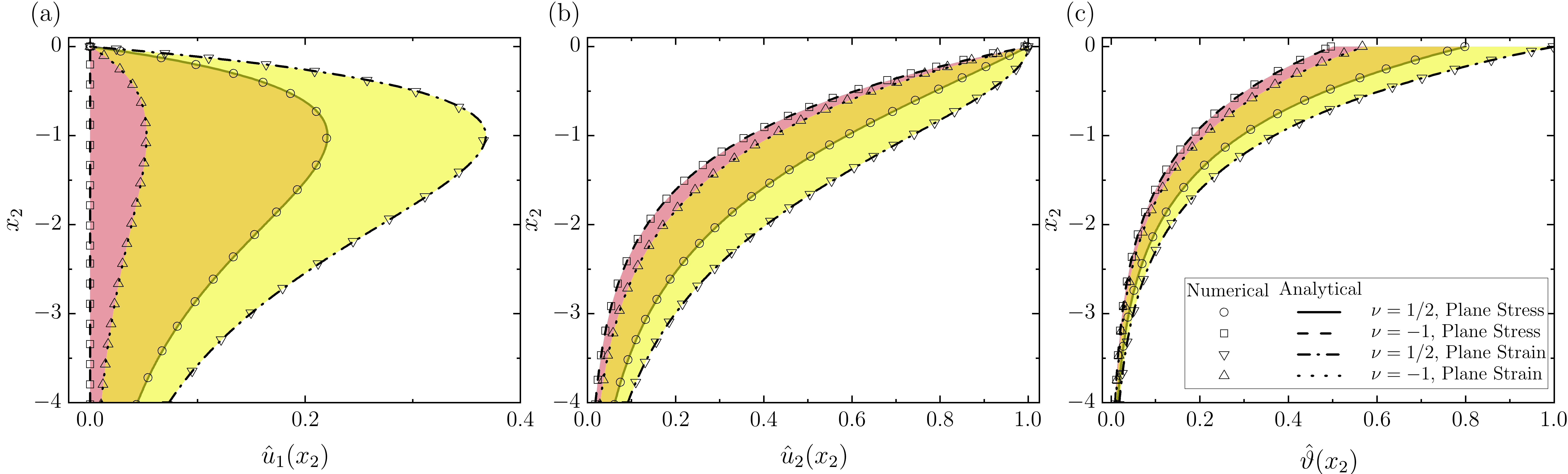}
    \caption{Displacement and microrotation amplitude functions along the thickness of the core for case A. The values of the respective amplitude function are shown in the abscissa as functions of coordinate along the core depth in the ordinate. Horizontal displacement field (a), Vertical displacement field (b), Microrotation field (c).}
    \label{fig:case0_distributions}
\end{figure}

We now proceed by solving the boundary value problem (BVP) for a half-infinite core lying at the bottom side of the Euclidean plane ($x_2\leq0$)---as seen in~\cref{fig:cases}-(a). We require decaying exponential terms by setting: $c_1=c_3$, $c_5=c_7$ and $c_6=c_8$ in~\cref{eq:sols}. The remaining constants are found by setting the interface to be normal strain free in the $x_1$ direction: $\varepsilon_{11}(x_1,0)=0$ and, by setting the interface displacement in the $x_2$ direction to match the deflection of the beam: $u_2(x_1,0)=w_0\sin(x_1)$. Furthermore, we must ensure that the medium is locally non-micropolar at the interface hence, the last integration constant will be found by matching microrotation and infinitesimal rotation at the interface ($x_2=0$), which is written as $\epsilon_{\alpha\beta}\partial_\alpha u_\beta=2\vartheta$. Resolving the boundary conditions yields the following forms of the stress functions:
\begin{subequations}
\begin{align}
    \Phi(x_\alpha)&= -w_0\frac{2\nu^*\left(1+g\right)+x_2\left(1-2\nu^*\right)}{1+2\nu^*}e^{x_2}\sin{x_1}\\
    \Psi(x_\alpha)&=w_0\frac{2\nu^*g}{1+2\nu^*}e^{x_2}\cos{x_1}.
\end{align}
\end{subequations}
We then calculate the normalised restoring normal and shear stresses and the resulting stiffness at the interface respectively as:
\begin{subequations}
    \begin{align}
    \sigma_{22}&=\frac{2\nu^*}{1+2\nu^*}w_0\sin x_1\\
    \sigma_{21}&=\frac{1}{1+2\nu^*}w_0\cos x_1\\
    k_n&=\frac{2\nu^*}{1+2\nu^*}\\
    k_s&=\frac{1}{1+2\nu^*}.
    \end{align}
\end{subequations}
Similarly, we also need to account for a restoring normalised couple stress and resulting rotational stiffness at the interface respectively as:
\begin{subequations}
    \begin{align}
    m_{23}&=g\frac{2\nu^*}{1+2\nu^*}w_0\cos x_1\\
    k_\vartheta&=g\frac{2\nu^*}{1+2\nu^*}=g k_n.
    \end{align}
\end{subequations}
Note that for the limit cases regarding wavelengths, the foundation would be rigid ($w_0\rightarrow\infty$) for $\ell\rightarrow0$ and infinitely soft ($w_0\rightarrow0$) for $\ell\rightarrow\infty$ in both linear and rotational terms, since there is no any other length-scale dependence for this particular case. That suggests that the dimensional restoring stress is inversely proportional to the length-scale of the applied disturbance. Furthermore, when the micropolar effects vanish i.e., $\ell_b\rightarrow0$ then the only contribution to the foundation stiffness will be through the linear stiffness terms which recovers the solution of~\cite{allen1969}.
The resulting displacement fields are then written using amplitude functions to split from the trigonometric terms as:
\begin{subequations}
    \begin{align}
        u_1(x_\alpha)&=w_0\hat{u_1}(x_2)\cos x_1=w_0\frac{(1-2\nu^*)x_2}{1+2\nu^*}e^{x_2}\cos x_1\\
        u_2(x_\alpha)&=w_0\hat{u_2}(x_2)\sin x_1=w_0\frac{1-2\nu^*+(1+2\nu^*)x_2}{1+2\nu^*}e^{x_2}\sin x_1\\
        \vartheta(x_\alpha)&=w_0\hat{\vartheta}(x_2)\cos x_1=w_0\frac{2\nu^*}{1+2\nu^*}e^{x_2}\cos x_1.
    \end{align}
\end{subequations}
The shapes of the amplitude functions, $\hat{u_1}$, $\hat{u_2}$, and $\hat{\vartheta}$ for the case of infinite core thickness are shown in~\cref{fig:case0_distributions} for all the acceptable values of the Poisson's ratio and for both plane stress and plane strain conditions. Note, that the range of the decaying displacement fields is approximately $5\ell$ through the depth of the core. Notice that, as a consequence of the separation of variables ansatz used, both the horizontal displacement and microrotation fields will have a $\pi/2$ phase difference to the vertical displacement field. Both plane strain and plane stress cases are shown for the complete range of thermodynamically acceptable Poisson's ratio values. For the horisontal displacement amplitude, we observe that it has a maximum value at $x_2=-1$ and then decays exponentially. For both vertical displacement and microrotation amplitudes, we observe an exponentially decaying behaviour. However, the vertical displacement will always exhibit a unit displacement amplitude at the boundary because of the boundary condition, contrary to the microrotation amplitude where the maximum value is found at the boundary and is depending on the Poisson's ratio. 

\subsection{Case B: Finite core foundation on rigid and deformed boundaries}
We proceed by solving the case of a finite core, which is fixed in one side and it is deformed with a sine-wave profile on the free surface---as seen in~\cref{fig:cases}-(b). We choose $\tilde{x}_2=x_2=0$ for the fixed side and $\tilde{x}_2=\tilde{c}=\ell c$ for the deformed side, where $\tilde{c}$ and $c$ is the dimensional and normalised thickness of the core respectively. The boundary conditions that make one side displacement free are given by $u_1(x_1,0)=u_2(x_1,0)=0$. For the deformed side, we use the same boundary conditions as for the infinite case. Namely, at $x_2=c$ we set $u_2=w_0\sin x_1$, and $\varepsilon_{11}(x_1,0)=0$. The $3^\text{rd}$ boundary condition needed at the lower side is chosen to match the rotations, \emph{i.e.} $\epsilon_{\alpha\beta}\partial_\alpha u_\beta=2\vartheta$, since the domain is assumed macroscopically fixed. 
The resulting integration constants for this case are:
\begin{subequations}\begin{align}
    c_1&=w_c\frac{4 c (g+1) \nu ^* \left[c
   \left(1-2 \nu ^*\right) \cosh c-(1+2 \nu ^*) \sinh c\right]}{\left(1+2 \nu
   ^*\right)^2 \left[\cosh (2 c)-1\right]-2 c^2 \left(1-2 \nu ^*\right)^2}\\
    c_3&=w_c\frac{2 c^2 \left(1-2 \nu ^*\right) 
   \left(1-2 g \nu ^*\right)\sinh c}{\left(1+2 \nu
   ^*\right)^2 \left[\cosh (2 c)-1\right]-2 c^2 \left(1-2 \nu ^*\right)^2}\\
    c_5&=w_c\frac{4 c^2 g \nu ^* \left(1-2 \nu ^*\right) \sinh
   c}{\left(1+2 \nu
   ^*\right)^2 \left[\cosh (2 c)-1\right]-2 c^2 \left(1-2 \nu ^*\right)^2}\\
    c_7&=w_c\frac{4 c g \nu ^* \left[(1+2 \nu ^*) \sinh c+c
   \left(2 \nu ^*-1\right) \cosh c+\right]}{\left(1+2 \nu
   ^*\right)^2 \left[\cosh (2 c)-1\right]-2 c^2 \left(1-2 \nu ^*\right)^2}.
   \end{align}
\end{subequations}
The resulting normal and shear stresses at the interface and are then written as
\begin{subequations}
    \begin{align}
    \sigma_{22}&=\frac{2c \nu^*    \left[(1+2 \nu^*) \sinh (2 c)-2c (1-2 \nu^*)\right]}{ (1+2 \nu^*)^2\left[ \cosh (2 c )-1\right]-2c^2 (1-2
   \nu^*)^2}w_c \sin x_1\\
   \sigma_{21}&=c\frac{(1+2\nu^*)\left[1-\cosh(2c)\right]+2(1-2\nu^*)^2c^2}{(1+2\nu^*)^2\left[1-\cosh(2c)\right]+2(1-2\nu^*)^2c^2} w_c\cos x_1
     \end{align}
\end{subequations}
 and, consequently, their respective dimensional stiffnesses are
 \begin{subequations}
    \begin{align}
    k_n'(c)&= \frac{2c \nu^*    \left[(1+2 \nu^*) \sinh (2 c)-2c (1-2 \nu^*)\right]}{ (1+2 \nu^*)^2\left[ \cosh (2 c )-1\right]-2c^2 (1-2
   \nu^*)^2}\\
   k_s'(c)&=c\frac{(1+2\nu^*)\left[1-\cosh(2c)\right]+2(1-2\nu^*)^2c^2}{(1+2\nu^*)^2\left[1-\cosh(2c)\right]+2(1-2\nu^*)^2c^2}.
    \end{align}
\end{subequations}
Similarly, the restoring couple stress is found to be
\begin{equation}
    m_{23}=\frac{2c \nu^*    \left[(1+2 \nu^*) \sinh (2 c)-2c (1-2 \nu^*)\right]}{ (1+2 \nu^*)^2\left[ \cosh (2 c )-1\right]-2c^2 (1-2
   \nu^*)^2}w_c \cos x_1,
\end{equation}
whereas the corresponding torsional stiffness would be
\begin{equation}
 k_\vartheta'(c)=g k_n'(c).
\end{equation}
The effects of the normalised core thickness on the stiffness are then investigated. The limit stiffness for thin cores and large wavelengths ($c\rightarrow0$), also known as elastic foundation assumption or Winkler's assumption, is calculated as well as the limit for negligible micropolar effects ($g\rightarrow0$ or $\ell_b\rightarrow0$) that is resembling classical elasticity.~\rv{Therefore, the classical elastic foundation solution (Cauchy formalism) can be recovered as a special case of the developed theory through the latter limit case:}
\begin{equation}
    \lim_{c \to 0} k_n'(c)=\nu^*,\;\;\lim_{c\to 0} k_s'(c)=0,\;\;\lim_{c \to 0} k_\phi'(c)=g\nu^*,\textrm{ and }\lim_{g \to 0} k_\phi'(c)=0.
\end{equation}

\begin{figure}[h!]
    \centering
    \includegraphics[width=\textwidth]{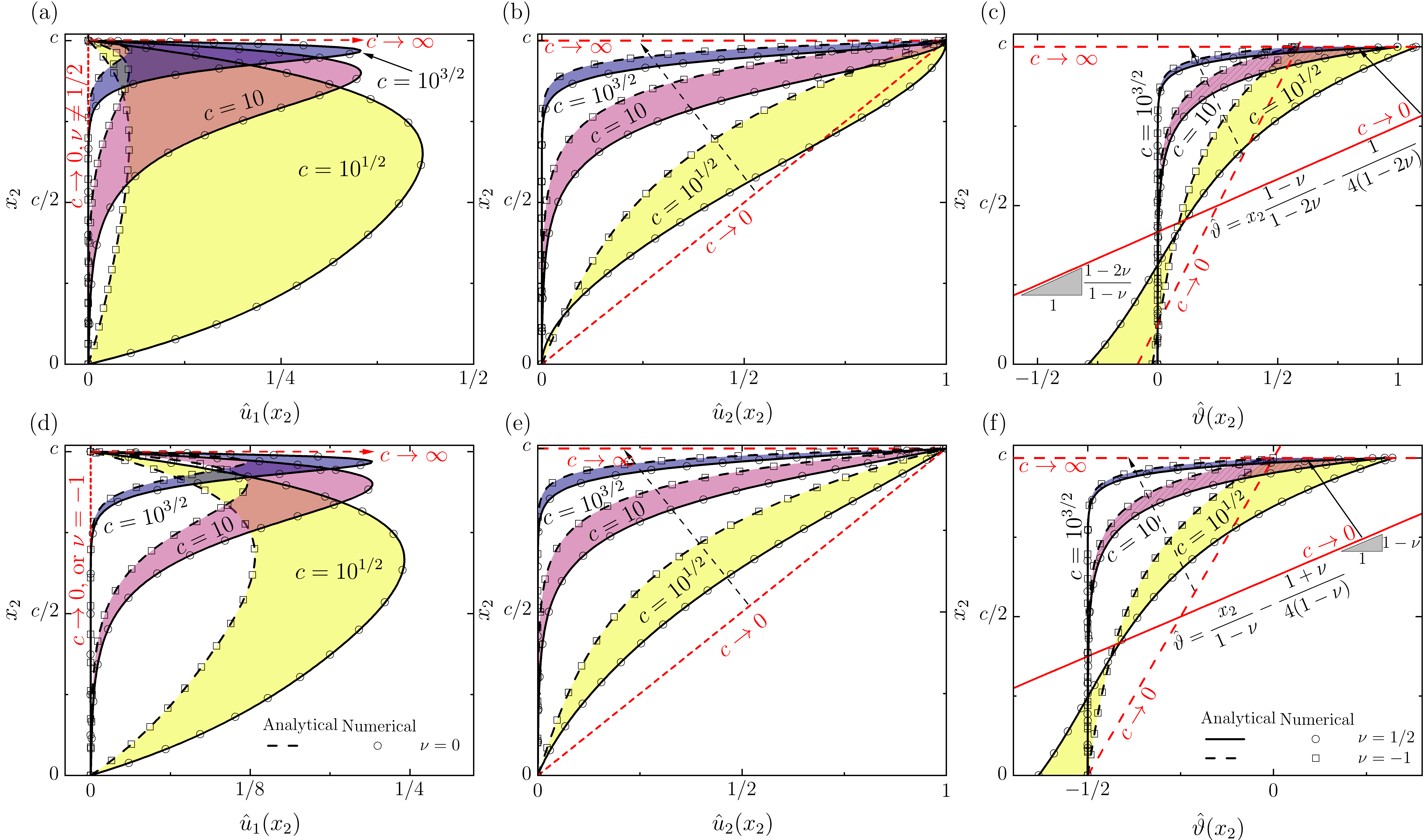}
    \caption{Displacement and microrotation amplitude functions along the thickness of the core for case B. The values of the respective amplitude are shown in the abscissa as functions of the coordinate along the core depth in the ordinate. Plane strain (a-c) and plane stress (d-f). Horizontal displacements (a, d), vertical displacements (b, e), microrotations (c, f).}
    \label{fig:case1_distributions}
\end{figure}

The displacement amplitudes are shown in~\cref{fig:case1_distributions}, where we present a range of normalised core thicknesses to show the effects on the different amplitudes. The amplitude of the horizontal displacement shows a maximum value that is positioned close to the centre of the core, for low core thicknesses, while the value of the maximum depends on the Poisson's ratio. As the relative thickness increases, the maximum's location moves towards the interface forming a pulse of constant amplitude and decreasing width in order to satisfy the zero boundary condition at the interface. Regarding the infinitely thin and infinitely thick core limits, the horisontal displacement will be zero at the former and an infinite pulse at the latter. The vertical displacement tends to be linear as the core's thickness approaches zero. As the thickness increases, the vertical displacement concentrates at the interface and for very thick cores it tends to a step shape. Vertical displacement's limit for infinitely thin core will be a linear distribution along the thickness, and a step distribution for infinitely thick core. The microrotation follows a similar trend to the vertical displacement. However, its value at the interface is depending on the value of the Poisson's ratio since the microrotation's boundary condition is such to balance the infinitesimal rotation. Therefore, microrotation's amplitude at the limit case for infinitely thin core will be a family of lines with slope and intercept depending on the Poisson's ratio as shown. Note, that when Poisson's ratio is exactly $1/2$ for plane strain, the zero thickness limit line will be a horisontal line at $x_2=c/2$. We use arrows to show the trend between the infinitessimal and infinite core thickness. In the microrotation amplitude, as the core thickens, all different lines will collapse at the horisontal pulse at $x_2=c$ which is independent of the Poisson's ratio. Lastly, we plot numerical results alongside the analytical lines by using markers and we observe a very good agreement.
\subsection{Case C: Finite core foundation on anti-symmetric boundaries}
For the case of anti-symmetrically deformed boundaries (\cref{fig:cases}-(c)), we choose the two boundaries to be at $x_2=\pm c/2$. The two boundaries are deformed in the same way, hence, at both boundaries, $x_2=\pm c/2$, we set the following conditions: $u_2=w_0\sin x_1$, $\varepsilon_{11}=0$ and $\epsilon_{\alpha\beta}\partial_\alpha u_\beta=2\vartheta$. For this set of boundary conditions, the constants $c_1=c_7=0$ vanish, hence the remaining integration constants are:
\begin{subequations}\begin{align}
    c_3&=w_c\frac{c^2\left(1-2  \nu ^*\right) \cosh \left(c/2\right)-4c (g+1)
   \nu ^* \sinh \left(c/2\right)}{(2\nu^*+1)\sinh c-(1-2\nu^*)c}\\
    c_5&=\frac{4 c g \nu ^* \sinh \left(c/2\right)}{(2\nu^*+1)\sinh c-(1-2\nu^*)c}
\end{align}\end{subequations}
The restoring stresses and their respective foundation stiffnesses at the interface $x_2=c/2$ then yield as:
\begin{subequations}
    \begin{align}
    \sigma_{22}&=\frac{4\nu^* c \sinh^2\left(c/2\right)}{\left(1+2\nu^*\right)\sinh c-\left(1-2\nu^*\right)c}w_c \sin x_1\\
    \sigma_{21}&=\frac{c \sinh c-c^2\left(1-2\nu^*\right)  \sinh^2\left(c/2\right)}{\left(1+2\nu^*\right)\sinh c-\left(1-2\nu^*\right)c}w_c \cos x_1\\
    k_n'\left(c\right)&=\frac{4\nu^* c \sinh^2\left(c/2\right)}{\sinh c \left(1+2\nu^*\right)-c\left(1-2\nu^*\right)}\\
    k_s'\left(c\right)&=\frac{c \sinh c-c^2\left(1-2\nu^*\right)  \sinh^2\left(c/2\right)}{\sinh c \left(1+2\nu^*\right)-c\left(1-2\nu^*\right)}.    
    \end{align}
\end{subequations}
Similarly, for the restoring couple stress and torsional stiffness, we obtain:
\begin{subequations}
    \begin{align}
    m_{23}&=\frac{4\nu^* c g \sinh^2\left(c/2\right)}{\left(1+2\nu^*\right)\sinh c-\left(1-2\nu^*\right)c}w_c \cos x_1\\
    k_\vartheta'\left(c\right)&=g k_n'(c).
    \end{align}
\end{subequations}
The limit stiffness for very thin cores is calculated (elastic foundation assumption) as well as the limit for negligible micropolar effects,~\rv{once again the classical solution is recovered}:
\begin{equation}
    \lim_{c \to 0} k_n'(c)=0,\;\; \lim_{c \to 0} k_s'(c)=0,\;\; \lim_{c \to 0} k_\vartheta'(c)=0,\textrm{ and }\lim_{\ell_b \to 0} k_\vartheta'(c)=0.
\end{equation}

\begin{figure}[htb!]
    \centering
    \includegraphics[width=\textwidth]{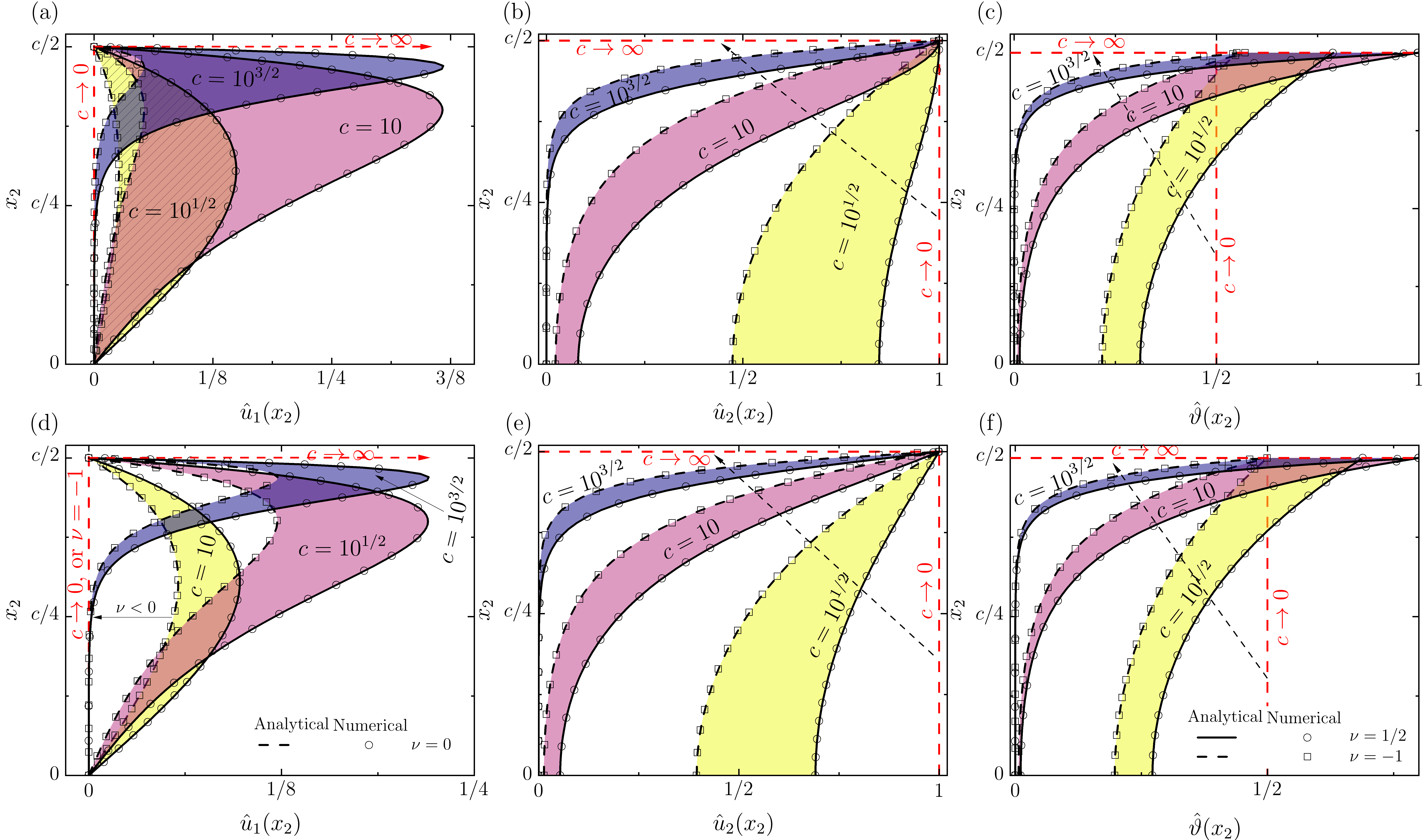}
    \caption{Displacement and microrotation amplitude functions along the thickness of the core for case C. The values of the respective amplitude are shown in the abscissa as functions of coordinate along the core depth in the ordinate. Plane strain (a-c) and plane stress (d-f). Horizontal displacements (a, d), vertical displacements (b, e), microrotations (c, f).}
    \label{fig:case2_distributions}
\end{figure}

The displacement amplitudes are drawn in~\cref{fig:case2_distributions}. We present a range of normalised core thicknesses to show the effects on the different amplitudes. Because of anti-symmetry, at this load case we expect the distributions to be evenly shaped and thus we present only the upper half of the core thickness. Regarding the horizontal displacement distributions, we observe a similar behaviour to case B because of the similarity of the boundary conditions at this field. However, for vertical displacement distribution, at the limit for very thin core, shows a constant value of the vertical displacement along the core since both boundaries are deformed in the same way. For very thick cores, the two boundaries are behaving independently and we recover a similar shape with case B. The microrotation is following a distribution similar to the vertical displacement with the value at the interface depending both on the Poisson's ratio and the thickness of the core giving larger amplitudes for thicker cores. 
Therefore, microrotation's amplitude at the limit case for infinitely thin core will be a vertical line that is dependent on the Poisson's ratio. Lastly, we plot numerical results alongside the analytical lines by using markers and we observe a very good agreement.
Finally, at the limit $c\rightarrow0$ the displacements at the two boundaries are equalized between them, hence the stresses at the limit case are zero.
\subsection{Case D: Finite core foundation on symmetric boundaries}
For the case of symmetrically deformed boundaries (\cref{fig:cases}-(d)), the two boundaries are deformed in the opposite way, hence, at $x_2=\pm c/2$ we set: $u_2=\pm\sin x_1$, $\varepsilon_{11}=0$ and $\epsilon_{\alpha\beta}\partial_\alpha u_\beta=2\vartheta$. 
For this load case, the vanishing integration constants are $c_3=c_5=0$, hence the remaining integration constants are:
\begin{subequations}
    \begin{align}
    c_1&=w_c \frac{  c^2 \left(1-2 \nu ^*\right)\sinh \left(c/2\right)-4c (g+1) \nu ^* \cosh
   \left(c/2\right)}{c(1-2\nu ^*)+(1+2\nu^*)\sinh
   c}\\
    c_7&=w_c\frac{4 c g \nu ^* \cosh \left(c/2\right)}{c(1-2\nu ^*)+(1+2\nu^*)\sinh
   c}.
    \end{align}
\end{subequations}
Hence, the restoring stresses at the interfaces and their respective stiffnesses are written as:
\begin{subequations}
    \begin{align}
    \sigma_{22}&=\frac{4\nu^* c \cosh^2\left(c/2\right)}{(1+2\nu^*)\sinh c+(1-2\nu^*)c} w_c \sin x_1\\
    \sigma_{21}&=\frac{c^2(1-2\nu^*)+c \sinh c}{(1+2\nu^*)\sinh c+(1-2\nu^*)c}w_c \cos x_1\\
    k_n'(c)&=\frac{4\nu^* c \cosh^2\left(c/2\right)}{(1+2\nu^*)\sinh c+(1-2\nu^*)c}\\
    k_s'(c)&=\frac{c^2(1-2\nu^*)+c \sinh c}{(1+2\nu^*)\sinh c+(1-2\nu^*)c}.
    \end{align}
\end{subequations}
Similarly, for the restoring couple stress we get:
\begin{subequations}
    \begin{align}
    m_{23}&=\frac{4\nu^* g c \cosh^2\left(c/2\right)}{(1+2\nu^*)\sinh c+(1-2\nu^*)c} w_c \cos x_1\\
    k_\vartheta'(c)&=g k_n'(c).
    \end{align}
\end{subequations}
\rv{Once again, the classical stiffness value for extremely thin cores is determined using the elastic foundation assumption, alongside the limit for negligible micropolar effects. This is seen as follows:}
\begin{equation}
    \lim_{c \to 0} k_n'(c)=2\nu^*,\;\; \lim_{c \to 0} k_s'(c)=0,\;\; \lim_{c \to 0} k_\vartheta'(c)=2g\nu^*,\textrm{ and }\lim_{\ell_b \to 0} k_\vartheta'(c)=0.\label{eq:limits}
\end{equation}

\begin{figure}[htb!]
    \centering
    \includegraphics[width=\textwidth]{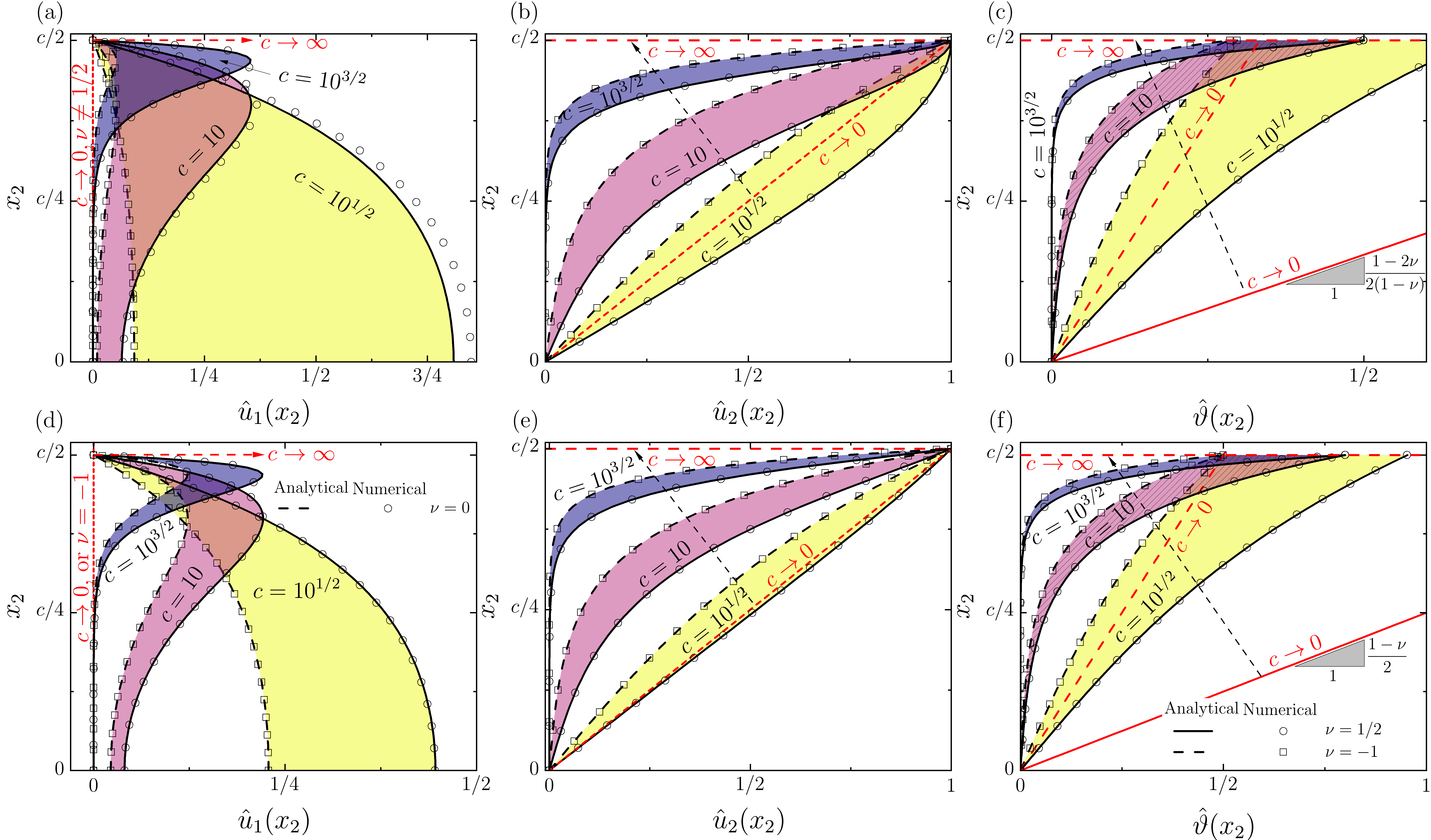}
    \caption{Displacement and microrotation amplitude functions along the thickness of the core for load case D. The values of the respective amplitude are shown in the abscissa as functions of the functions of coordinate along the core depth in the ordinate. Plane strain (a-c) and plane stress (d-f). Horizontal displacements (a, d), vertical displacements (b, e), microrotations (c, f).}
    \label{fig:case3_distributions}
\end{figure}

\begin{figure}
    \centering
    \includegraphics[width=\textwidth]{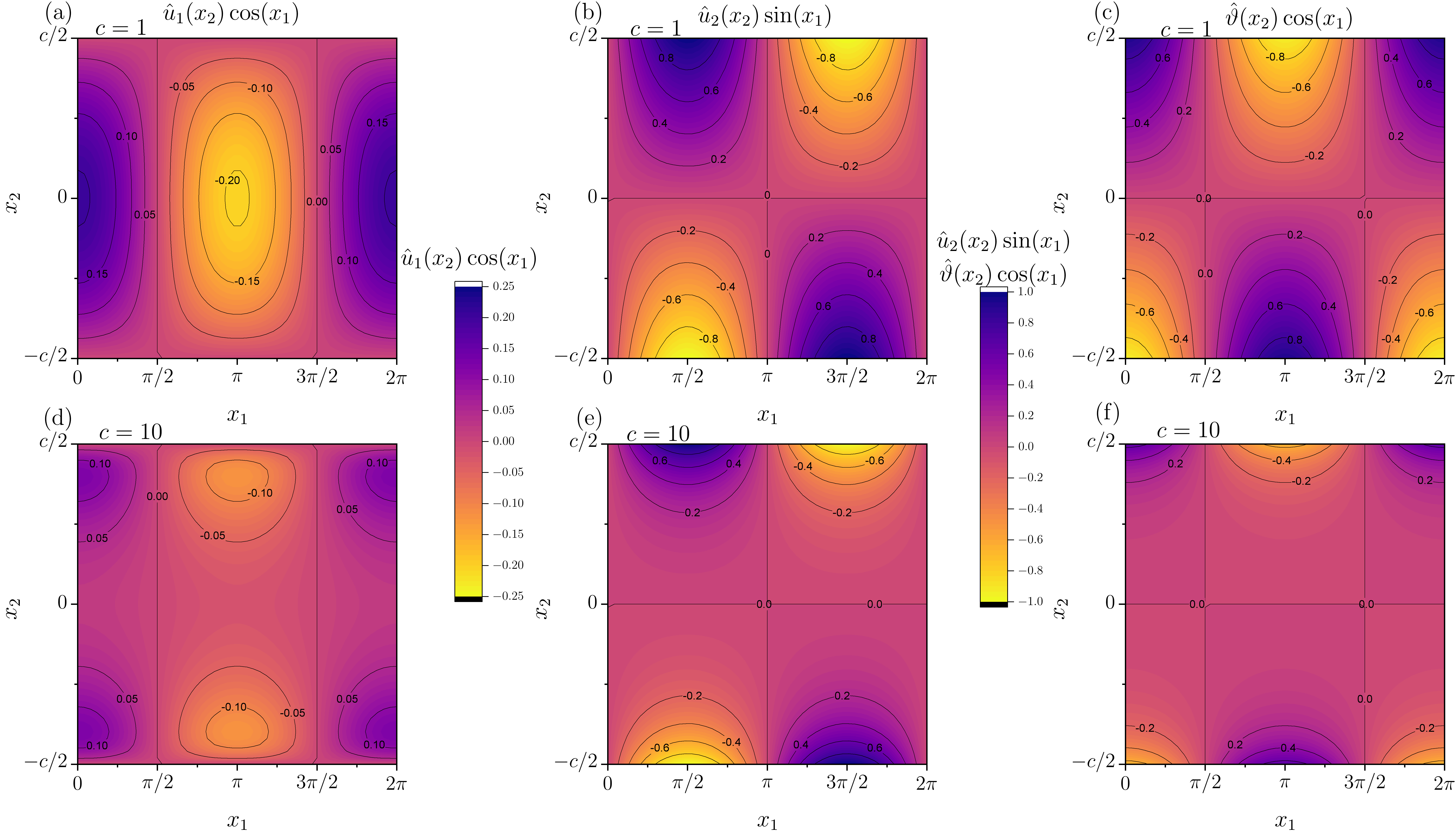}
    \caption{Contours of normalised displacement and microrotation distributions for $c=1$ (a-c) and $c=10$ (d-f) for the antisymmetric boundary condition case (case d). Horizontal displacements are shown in plots a and d, vertical displacements are shown in plots b and e, and microrotations are shown in plots c and f. The boundary effects can be observed merged for low values of normalised core thickness ($c=1$) resulting to linearly varying amplitudes for vertical displacement and microrotation. For high values of normalised core thickness ($c=10$), the boundary effects are localised at the two boundaries resulting to two separate deformed regions of the bulk core.}
    \label{fig:contours}
\end{figure}

\rv{As in the previous cases, the displacement amplitudes for the symmetric load case are here depicted in~\cref{fig:case3_distributions}. We once again illustrate various normalized core thicknesses to demonstrate their effects on distinct amplitudes. Due to symmetry, we also anticipate the distributions in this case to be asymmetric and, therefore, present solely the upper half of the core thickness.} All three field distributions present great similarity to case B with a few key differences. The slope of the vertical displacement at the limit case is double, since displacement is applied from both sides, and the microrotation is distributed symmetrically. It is also observed that the vertical displacement's slope is double than that of case B, since the total applied displacement magnitude is double due to added contributions from top and bottom boundaries. The amplitude of the microrotation at the limit case for infinitely thin core will be a family of lines passing through the origin, with slope depending on the Poisson's ratio as shown. Lastly, we plot numerical results alongside the analytical lines by using markers and we observe a very good agreement. We also present the complete normalised displacement and microrotation distributions (\cref{fig:contours}) for two representative cases $c=1$ and $c=10$ and assuming a material with $\nu=0$ (or $\nu^*=1$). Through~\cref{fig:case3_distributions,fig:contours} the merging and separation of the two boundary effects can be observed. Specifically, when $c=10$ there is enough bulk for both localised boundary effects to fully develop. However, when $c=1$ the bulk is relatively thin, hence the distributions converge to linear through the thickness. 
\subsection{Stiffness response of the core}
\begin{figure}
    \centering
    \includegraphics[width=\textwidth]{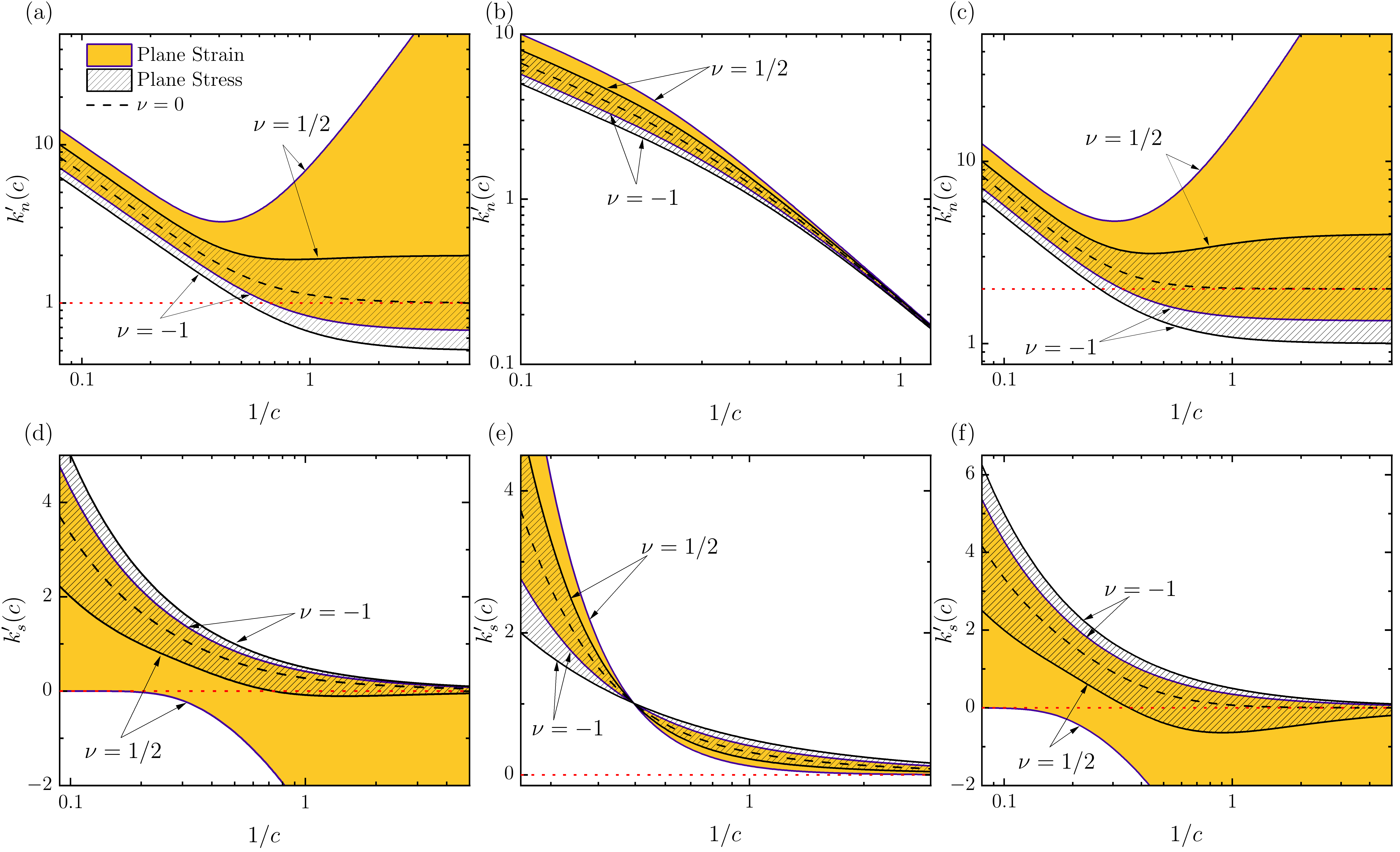}
    \caption{Restoring stiffness as functions of the normalised core thicknes for load case b (plots a and d), load case c (plots b and e) and load case d (plots c and f). We present normal restoring stiffness curves (a-c) as well as shear restoring stiffness curves (d-e). It is observed that for thin enough cores the restoring normal stiffness is approaching asymptotically a constant value that is a function only of the Poisson's ratio. The shear stiffness is always approaching zero asymptotically. However, the best convergence speed is achieved for $\nu=0$ while for high values of the Poisson's ratio, a negative restoring shear stress can be observed.}
    \label{fig:stiffness}
\end{figure}
Subsequently, the stiffness behaviour of the bulk core needs to be assessed for all the cases modelled. The complete normal and shear stiffness response of each case is displayed in~\cref{fig:stiffness} as a function of the inverse normalised core thickness. Cases B and D converge to specific values of normal stiffness for thin cores, where converging values of case D are generally double of those of case B. In case C, the normal stiffness always decreases with the thickness, leading to zero normal stiffness at the thin core limit. Regarding the shear stiffness, it is always vanishing small, asymptotically zero, as the core gets thinner. It is also noted that its convergence to zero is faster when $\nu=0$ and that it may take negative values for $\nu>0$. By comparing the normal stiffness behaviour and the through thickness vertical displacements shown in~\cref{fig:case1_distributions,fig:case2_distributions,fig:case3_distributions}, we may conclude that two dominant behaviours are present: (i) constant stiffness for small core thicknesses corresponding to a linear displacement distribution along the thickness of the core, and (ii) an exponentially increasing stiffness as the core thickness increases that corresponds to a step-like shaped displacement distribution. Such step-like displacement distribution which is also observed for the infinite core case in~\cref{fig:case0_distributions}-(b). Thereafter, we may attribute the change of the core's behaviour to merging effects between the two localised deformation zones of the two boundaries. Namely, at the thin core limits, and for all three finite core cases that have been studied, the deformation zone cannot develop fully, leading to the following situations: (i) a linear displacement distribution from rigid to fully developed displacement resulting to constant through thickness stresses (case B), (ii) a constant fully developed displacement distribution through the thickness, that results in zero stresses (case C), and (iii) a linear displacement distribution from negative fully developed displacement to positive fully developed displacement, resulting to constant through thickness stresses. Furthermore, the stresses at case D are double of those found in case B since the displacement at the top and bottom are superimposed. It is worthy of note that, in the context of adhesives, such effects contribute to the fact that thick bondlines are more prone to failure due to delamination contrary to cohesive failure. More specifically, delamination is by far the most commonly observed failure mode in sandwich type structures and stiff adhesive joints with thick bondlines, where the stresses are highly concentrated at the corners because of the geometry of the boundary. Therefore, a narrow deformation zone that corresponds to large values of $c$---thick cores and low disturbance wavelengths---will contribute to directing a crack from the centre of the core towards the boundaries.

\section{Double Cantilever Beam configuration (DCB) on Micropolar Foundation}
\begin{figure}
    \centering
    \includegraphics[width=0.72\textwidth]{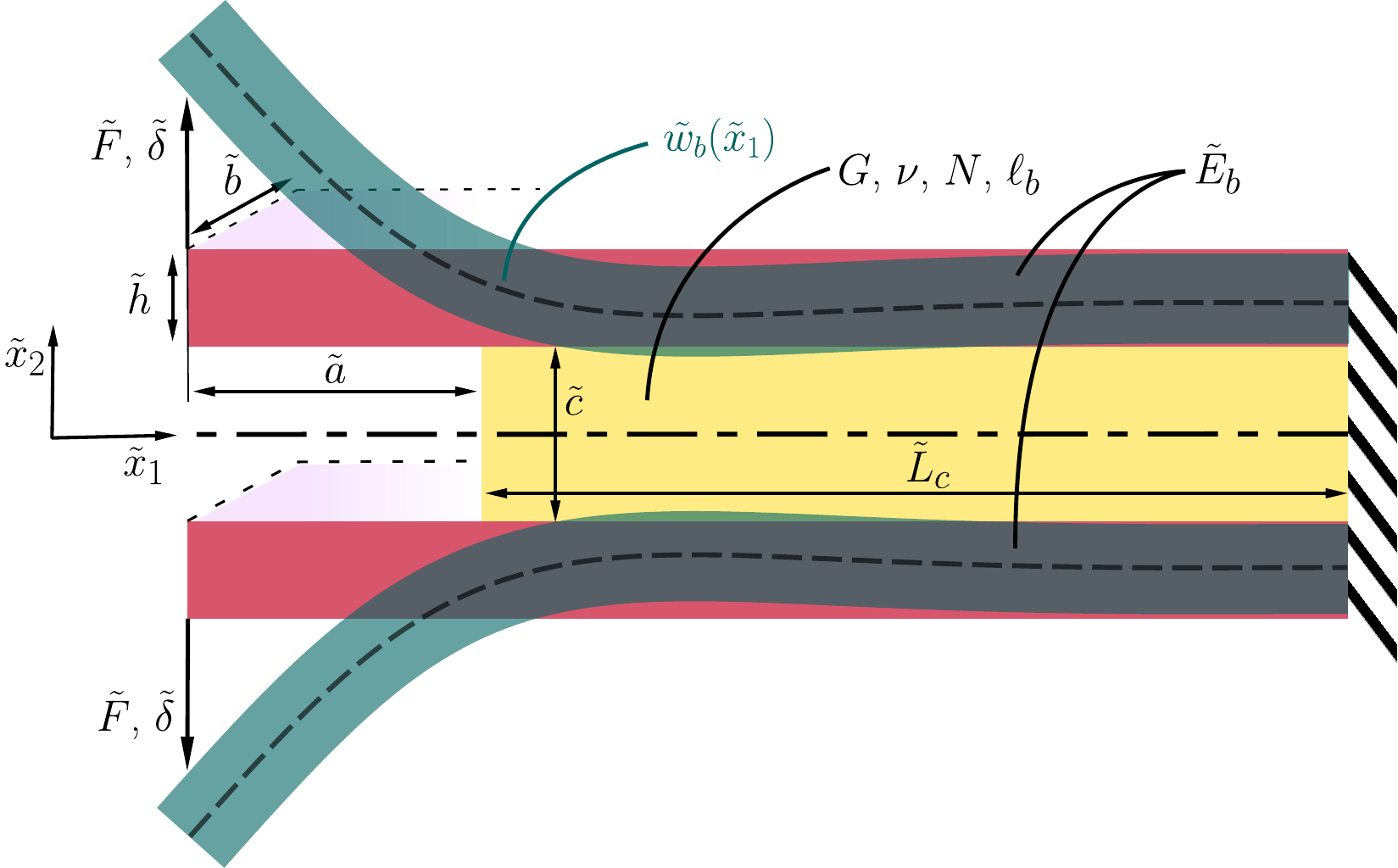}
    \caption{Schematic of the double cantilever beam configuration. Two Euler - Bernoulli beams of thickness $\tilde{h}$, length $\tilde{L}_\text{tot}$ and stiffness $\tilde{E}_b$ are bonded to a micropolar elastic core of thickness $\tilde{c}$, length $\tilde{L}_\text{c}$ and material properties $G$, $\nu$, $\ell_b$ and $N$. Both core and beams have an out of plane width $\tilde{b}$. The beams are bonded such that they extend at the left side creating a precrack of length $\tilde{a}$ and are assumed fixed at the right boundary of the bonded part. The free ends of the beams are then equaly displaced by $\tilde\delta$ that is balanced by a shear force $\tilde{F}$.}
    \label{fig:dcb}
\end{figure}

The aforementioned findings pertaining to the stiffness response of bulk micropolar cores, across the diverse loading configurations under consideration, afford us the capability to employ our developed theoretical framework for addressing certain aspects of fracture problems in heterogeneous media. We choose the double cantilever beam (DCB) configuration as a representative configuration for mode I fracture problems. The DCB configuration has been extensively studied, and associates the fracture process to a characteristic length, namely the fracture process zone, thus providing explicit knowledge of the stress distribution at the crack front. The DCB configuration is composed of two beams of thickness $\tilde{h}$ that are bonded with a core of thickness $\tilde{c}$. The total length $\tilde{L}_\text{tot}$ of the configuration is split into free length $\tilde{a}$ and bonded length $\tilde{L}_c=\tilde{L}_\text{tot}-\tilde{a}$. The free length is the length of the pre-crack and the bonded length is the length of the bondline. A detailed schematic of the DCB configuration is shown in~\cref{fig:dcb}. In this case, the beams are composed of 
a classically elastic material, whereas the core is assumed to be micropolar. 
\subsection{Euler-Bernoulli Beam theory formulation}
The solution of the DCB is obtained by assuming that the foundation is confined between two beams, thus reflecting case D for symmetrically loaded core. We also assume that the core is sufficiently thin and the characteristic length of the disturbance is sufficiently large to consider the stiffness value of the limit case. Since any response can be decomposed in harmonic terms, we may generally write the restoring stresses to be proportional to the kinematic variables as follows:
\begin{subequations}
    \begin{align}
    \sigma_{22}(x_1)&=k_n'(c) w(x_1),\\
    \sigma_{21}(x_1)&=k_s'(c) \partial_1 w(x_1),\\
    m_{23}(x_1)&=k_\vartheta'(c) \partial_1 w(x_1).    
    \end{align}\label{eq:stiffness}
\end{subequations}
This is consistent with~\cref{eq:stiffness_def}, where $w(x_1)\equiv \tilde{w}_b(\tilde{x}_1)/\tilde{c}$ is an arbitrary deflection curve normalised by the thickness of the elastic core. It is considered that the elastic beams are supported by linear spring and a rotational spring foundations, $k_n'$ and $k_\vartheta'$ respectively, at the limit case when $c\rightarrow0$---this is obtained from case D. The shear component does not offer a contribution, since in the limit case this this is strictly zero, \emph{i.e.} $k_s'\rightarrow0$. According to Euler-Bernoulli beam theory, the moment equilibrium equation for a beam subjected to a distributed force and moment loading conditions, reads as follows in its dimensionless form:
\begin{equation}
    \partial_1^2M-b \partial_1 m_{23}+b\sigma_{22}=0
\end{equation}
where $M\equiv \tilde{M}/(2G\ell^3)= E_b b c h^3 \partial_1^2 w(x)/12$ is the normalised moment and $E_b\equiv \tilde{E}_b/(2G)$, $b\equiv\tilde{b}/\ell$ and $h\equiv\tilde{h}/\ell$ are, respectively, the normalised stiffness, out of plane width, and thickness of the beam. By using the results from~\cref{eq:stiffness,eq:limits} we acquire the governing equation for the DCB configuration as:
\begin{equation}
    \partial_1^4 w-24\frac{g\nu^*}{E_b ch^3}\partial_1^2 w+24\frac{\nu^*}{E_b ch^3}w=0.\label{eq:BVP}
\end{equation}

From~\cref{eq:BVP}, it can be seen that a natural length scale defined as $\tilde\lambda^4\equiv E_b\tilde{c}\tilde{h}^3/(6\nu^*)$ arises. This is known as the process zone length. We also define its dimensionless version as $\lambda^4\equiv\tilde{\lambda}^4/\ell^4=E_b ch^3/(6\nu^*)$. When $\lambda=1$, then the wavelength of the disturbance matches the length of the process zone and the expressions given by~\cite{kanninen1973} and~\citep{Athanasiadis2021} can be recovered. Also, for $\lambda=1$ the BVP of~\cref{eq:BVP} is automatically normalised. Hence, we rewrite~\cref{eq:BVP} as:
\begin{equation}
    \partial_1^4w-2\xi\partial_1^2w+4w=0,\label{eq:BVP2}
\end{equation}
where $\xi=2g=(2\ell_b/\ell)^2=(2\ell_b/\tilde\lambda)^2>0$ is a modulator of the process zone that relates the material property associated to the micropolar length-scale $\ell_b$ to the length-scale of the disturbance $\ell$ of the configuration. More specifically, $\xi$ is a dimensionless parameter expressing the intensity of the rotational spring compared to the length of the fracture process zone. It is observed that when the micropolar behaviour of the core is ignored, then $\xi=0$ and the BVP falls back to the usual elastic foundation formulation. Also, the BVP of~\cref{eq:BVP2} becomes critical when $\xi=2$, implying that $\ell_b=\tilde\lambda/\sqrt{2}$. However, given that we have assumed that $\tilde{c}\ll\tilde\lambda$ and since $\ell_b$ is considered a property related to the microstructure of the micropolar core, we never expect to reach this condition. Hence, we will present solutions of the BVP for $0\leq\xi<2$. 

The complete problem consists of two regions. The beam is free (cantilever) at the first region at a length $a\equiv\tilde{a}/\lambda$ and supported by the foundation at the second region which is considered of infinite length. We set the origin where the bonded part begins. Then the beam is free for $x_1\leq0$ and bonded for $x_1>0$. Regarding the boundary conditions, we need to solve for eight constants. We consider the free end of the beam to be displaced by $\delta\equiv w(-a)=\tilde\delta/\tilde{c}$ and that the end is free of an applied moment, \emph{i.e.} $w''(-a)=0$. We require $\mathcal{C}^3$ continuity at $x_1=0$ to find four of the constants. The remaining two constants are calculated, by considering decaying exponential terms, which implies that the beam rests when $x_1\rightarrow\infty$. Therefore, the solution then reads:
\begin{equation}
  w(x_1)=\delta
  \begin{dcases}
  \frac{\hat\xi_++a(1+\xi)-(2a\hat\xi_++1)x_1+ax_1^2+x_1^3/3}{\hat{\xi}_++2a\hat\xi^2_++2a^2\hat\xi_++2a^3/3}, &x_1\leq0\\
  \frac{\hat{\xi}_-(\hat{\xi}_++a+a\xi)\cos(\hat\xi_-x_1)+\left[\xi/2-a(1-\xi)\hat\xi_+\right]\sin(\hat\xi_-x_1)}{\hat{\xi}_-(\hat{\xi}_++2a\hat\xi_+^2+2a^2\hat\xi_++2a^3/3)}e^{-\hat{\xi}_+x_1}, &x_1>0
  \end{dcases}\label{eq:EBsolution}
\end{equation}
where $\hat\xi_\pm=\sqrt{1\pm\xi/2}$. Moreover, the shear force acting at the free part arises as reaction at the free end $F\equiv\tilde{F}/(2G\ell^2)=b\nu^*\partial_1^3w/2$. Then, the normalized compliance of the system is written as follows:
\begin{equation}
    C\equiv\frac{2G\ell^2}{\tilde{c}}\tilde{C}=\frac{2}{b\nu^*}\frac{\delta}{\partial_1^3w(-a)}=\frac{1}{b\nu^*}\left(\hat\xi_++2a\hat\xi_+^2+2a^2\hat\xi_++\frac{2}{3}a^3\right),
\end{equation}
where $\tilde{C}=\tilde{\delta}/\tilde{F}(-\tilde{a})$ is the dimensional compliance of the system. Therefore, it is worthy of mention that this expression suggests that the system becomes more compliant as the micropolar contribution intensifies. 

\subsection{Numerical Implementation and Validation}
\begin{figure}
    \centering
    \includegraphics[width=\textwidth]{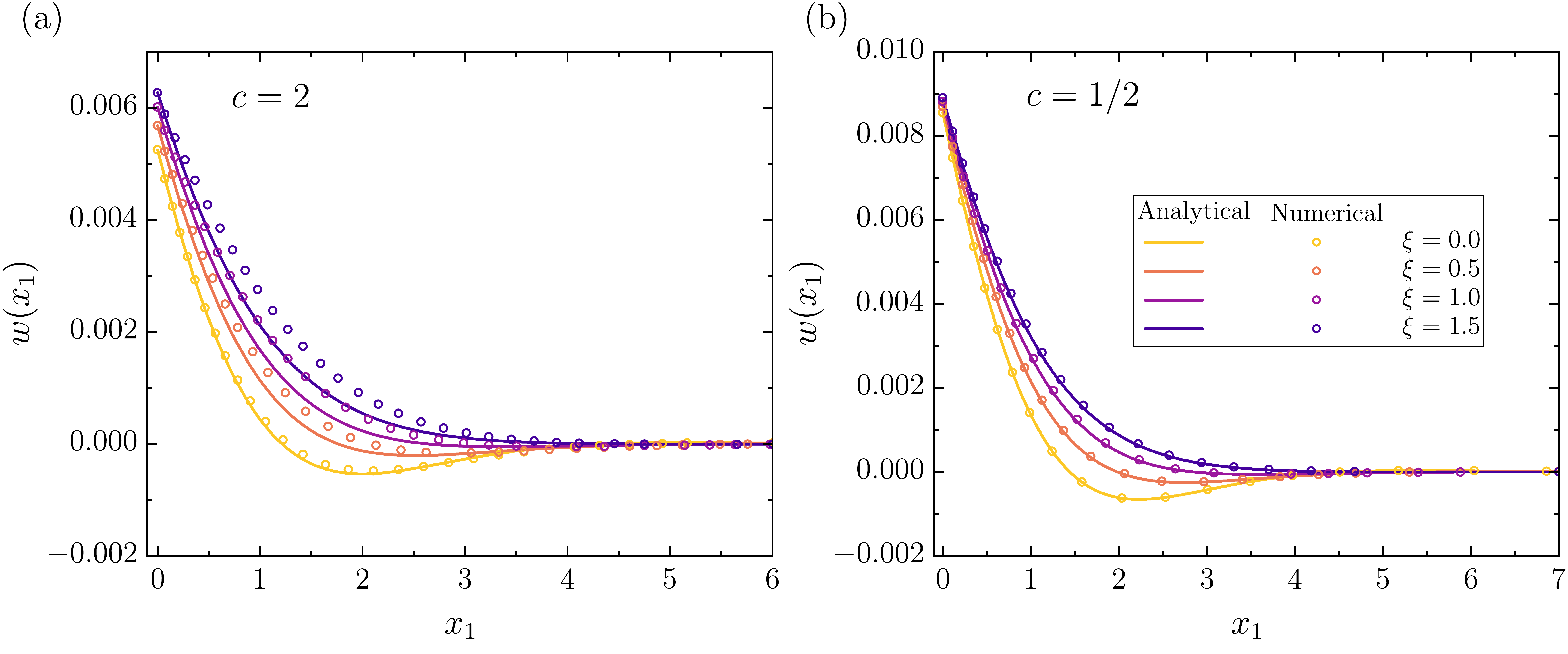}
    \caption{Justification of the accuracy of the analytical model against numerical solutions and parametric study of the rotational stiffness effects for the DCB configuration. When $c<1$ an acceptable accuracy is observed. The shape of the process zone is modulated through the parameter $\xi$ and the compression zone shortens as micropolar effects intensify.}
    \label{fig:deflections}
\end{figure}

Given the definitions from~\cref{sec2.1}, we may write the strain energy of the micropolar solid as follows:
\begin{equation}
    \mathcal{E}=\frac{1}{2}\int_\Omega\left(\tilde\sigma_{\alpha\beta}\tilde\varepsilon_{\alpha\beta}+\tilde{m}_\alpha\tilde\vartheta_{,\alpha}\right)\mathrm{d}\Omega.
\end{equation}
In order to identify the extrema of this functional, we take the first variation of the strain energy, thus yielding the weak form of the two dimensional micropolar theory of elasticity:
\begin{equation}
    \int_\Omega \left(\tilde\sigma_{\alpha\beta}\delta\tilde\varepsilon_{\alpha\beta}+\tilde{m}_\alpha\delta\tilde\vartheta_{,\alpha}\right)\mathrm{d}\Omega=0,
\end{equation}
where $\delta\tilde\varepsilon_{\alpha\beta}$ and $\delta\tilde\vartheta_{,\alpha}$ stand for the first variations of the strains and microcurvatures with respect to displacements and microrotation fields, respectively. Observe that the strains are functions of both displacements and microrotations, hence by definition the variation of the strain will contain terms accounting for both dependences. The weak form is implemented in COMSOL Multiphysics through the \emph{Weak Form PDE} physics interface using the \texttt{test} operator, similarly to other implementations developed in the micropolar context~\citep{suh2020}. The beams confining the micropolar core are modelled by using standard beam type elements that follow the Euler--Bernoulli formulation. These elements are quadratic in the rotation field and cubic in the displacement fields by default. Hence, the solutions are up to $\mathcal{C}^2$ continuous. With this consideration, we substitute the cantilever part of the beam $-a<x_1<0$ with appropriate displacement and rotational boundary conditions at $\tilde{x}_1=0$ that are extracted from~\cref{eq:EBsolution} evaluated at $\tilde{x}_1=0$. At the interface, we need to maintain the matching of displacements and loads between the two different components of our model. Therefore, the displacement fields and the infinitesimal rotation of the beam module are supplied as Dirichlet type boundary conditions at the edges of the micropolar core's displacements and microrotation respectively, where the former is ensuring the micropolar core to locally behave as a classical elastic solid at the edges. Regarding the matching of loads, we consider the beams to be loaded with distributed loads and moments opposite to the surface tractions produced by the micropolar solid, which are expressed as follows:
\begin{subequations}
\begin{align}
    \tilde{F}_\alpha&=-\tilde\sigma_{\beta\alpha}n_\beta\\
    \tilde{m}&=-\tilde{m}_\beta n_\beta.
\end{align}    
\end{subequations}
At last, we fix the other ends to ensure numerical stability. We demonstrate the validity of the thin core assumption as well as the effects of micropolar intensity by plotting~\cref{eq:EBsolution} alongside to numerical results produced by our \emph{in situ} implementation of the DCB configuration in~\cref{fig:deflections}. We use four representative values for the modulator $\xi=\{0.0, 0.5, 1.0, 1.5\}$, whereas the first case can also serve as the reference configuration for the absence of micropolar effects. \rv{Therefore, this falls back to the classical elastic foundation solution}. The deflection curves are presented for $c=\{2,1/2\}$, thus assessing the accuracy of the thin core assumption. Furthermore, for this demonstration we keep $\delta=0.01$ and the resulting normalised crack length is $a=\{0.6,0.15\}$ respectively to the chosen values of $c$. Regarding the material properties of this case, we have chosen $E_b=120$ and $\nu=0$.
\subsection{Applications to Foams and Regular Honeycombs}
\begin{figure}
    \centering
    \includegraphics[width=0.6\textwidth]{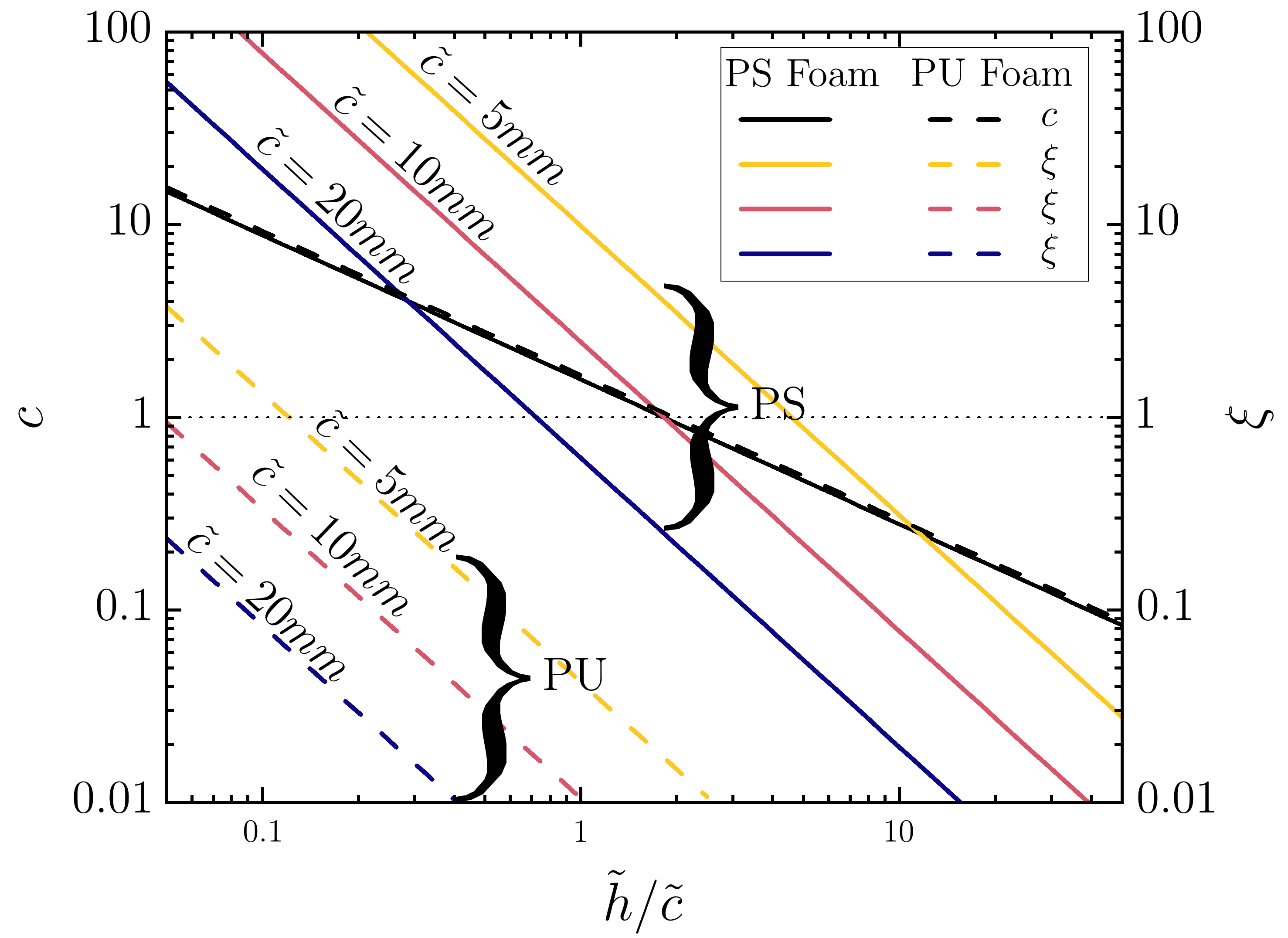}
    \caption{Investigation of the applicability of the theory for polystyrene (PS) and polyurethane (PU) foam cores under the DCB configuration. The significance of micropolar effects ($\xi$) and the normalised core thickness ($c$) are plotted against the thickness ratio of the constituents ($\tilde{h}/\tilde{c}$).}
    \label{fig:application}
\end{figure}

Based on our analysis, we attempt to quantify the significance of micropolar effects in the DCB configuration for cases where the core is composed of foams that are known to exhibit micropolar behaviour. We present a parametric study for the DCB configuration with Poly-Styrene (PS) and Poly-Urethane (PU) foams as core materials. To achieve that, the values of the parameter $\xi$ will be investigated. Following our definitions, we may write the following scaling laws for the parameter capturing the micropolar effect:
\begin{equation}
    \xi=\left(\frac{2\ell_b}{\tilde{\lambda}}\right)^2=\left(\frac{\tilde{h}}{\tilde{c}}\right)^{-3/2}\left(\frac{6\nu^*}{E_b}\right)^{1/2}\left(\frac{2\ell_b}{\tilde{c}}\right)^2.
\end{equation}
At the same time, we must also consider the relative core size, which scales as
\begin{equation}
    c=\frac{\tilde{c}}{\tilde{\lambda}}=\left(\frac{\tilde{h}}{\tilde{c}}\right)^{-3/4}\left(\frac{6\nu^*}{E_b}\right)^{1/4},
\end{equation}
in order to consider the validity of the Winkler assumption. However, a constrain of $c<1$ needs to be set in order to maintain the accuracy of the limit case, which constrains the values of $\xi$ accordingly:
\begin{equation}
    \xi<\left(\frac{2\ell_b}{\tilde{c}}\right)^2=\xi_\text{max}.
\end{equation}
This last inequality gives an estimation of the maximum possible contribution of the micropolar effects on the DCB configuration and relates a material parameter, \emph{i.e} the characteristic length for bending, to the overall geometry of the problem,\emph{i.e} the core thickness. This is required for maintaining the necessary assumptions of Winkler for elastic foundations. However, the characteristic length for bending is always related to some length of the microstructure. In the case of regular honeycombs, the characteristic length for bending at first order approximation for thin cell walls is $\ell_{b\text{,hex}}=L/(4\sqrt{3})<L$, where $L$ is the length of the cell wall~\citep{wang1998}. The core thickness for a cellular structure will always be an integer multiple of the unit cell size: $\tilde{c}\sim n L$ resulting to $\xi_\text{max}\sim 1/n^2$. For the extreme case of $n=1$, $\tilde{c}=2L$ and $\xi_\text{max}=1/48$, which restricts the micropolar contribution to a very low maximum that is negligible under Winkler's approximations. 
On the other hand, for the case of foams the characteristic length for bending may be spanning multiple unit cells ($\ell_{b\text{,foam}}>L$), thus allowing configurations with $\tilde{c}<2\ell_b$. This results into DCB configurations with sufficiently high values of $\xi_\text{max}$. For the case of Poly-Styrene foam (PS)~\citep{lakes1983}, $\ell_{b,PS}=5mm$ while the grain size is $L\approx1mm$. Thereafter, a configuration with $\tilde{c}=10mm$ will result to $\xi_\text{max}=1$. The exact value of $\xi$ will further depend on the rest of the parameters $\tilde{h}$, $\nu^*$, and the respective stiffness of the constituents. Similarly, for Poly-Urethane foam (PU)~\citep{lakes1986}, $\ell_{b,PU}=0.327mm$ with grain size $L\approx0.1mm$, hence meaning that $\tilde{c}=0.92mm$ would result to $\xi_\text{max}=0.5$. The trends for PS and PU foams are visualised in~\cref{fig:application}, where we plot the parameters $c$ and $\xi$ while varying the aspect ratio of the configuration ($\tilde{h}/\tilde{c}$) for fixed stiffness ratio and Poisson's ratio $E_b=10$ and $\nu^*=1$ respectively. It is seeing from~\cref{fig:application} that for acceptable accuracy a condition $c<1$ is needed, hence $\tilde{h}>2\tilde{c}$. On the other hand, to observe significant micropolar effects a condition $\xi>0.1$ is needed, leading to two competing quantities. Additionally, $\xi$ depends also on the nominal thickness of the core $\tilde{c}$, resulting in different trends that are scale dependent. This effect is born out of the fact that the micropolar characteristic length is always linked to the main length-scale of the microstructure, which here is constant.
\section{Conclusions}
In this study, we presented an initial attempt to apply micropolar elasticity theory to investigate a variety of issues related to heterogeneous foundations. Essentially, our theory extends Winkler's elastic foundation theory by incorporating additional internal degrees of freedom, specifically microrotations. This extension allows for the consideration of more complex structural configurations within heterogeneous materials. We developed a comprehensive theory of elastic foundations by examining the impact of microrotations when a system is subjected to various load configurations that could occur in practical scenarios. The primary outcome of our investigation was the derivation of analytical expressions for normal, shear, and rotational stiffness. Our objective was to provide a nuanced understanding of the responses associated with these stiffness parameters in more complex materials. This theoretical framework not only contributes to the fundamental knowledge base of micropolar elasticity but also lays the groundwork for potential practical applications. We hope that the insights gained from our analysis, regarding the behaviour of complex elastic media, may set the stage for further research, emphasising the significance of micropolar elasticity and elastic foundations in general.

We also illustrated the process of obtaining closed-form solutions that describe the deformation of a micropolar elastic solid confined between two Euler-Bernoulli beams, the DCB configuration. Our selected normalisation brought attention to critical parameters governing the behaviour of such a confined system. The micropolar characteristic length $\ell_b$ was shown to relate to the length of the fracture process zone, defining a modulator ($\xi$) for this zone. The modulator alters the deflection profile by reducing both the area of the core loaded by compressive stress and the intensity of compressive stresses.~\rv{When $\xi=0$, the solutions for the classical Cauchy continuum are recovered. Hence, it is essential that the modulator is considered during the analysis of confined media that are micropolar.} In the context of the DCB configuration, we validated the resulting deflection profiles and identified the regimes where the thin core assumption holds by comparing our solution to our custom implementation of micropolar elasticity in the COMSOL Multiphysics platform. Additionally, we demonstrated how the intensity of micropolar effects contributes to an increase in the total compliance of the system.

We also investigated the importance of considering the relative core thickness in the design process to classify a confined system as either having a thin core ($c<1$) or a thick core ($c>1$). We justified how the relative core thickness transitions the system from linear through-thickness displacement to step-like through-thickness displacement and discussed the effects of such behaviour on the restoring stiffness applied from the core to the confining beams. Additionally, we expanded the existing theory to encompass the equivalent microrotational field across the thickness and the corresponding rotational restoring stiffness experienced by the beam as a rotation. This newly discovered (moment) contribution is proportionate to both the normal (force) stiffness and the square of the micropolar characteristic length for bending $\ell_b$, which defines the intensity of the couple stress balancing microrotation gradients.

Finally, we derived a scaling law that relates our parameters to the dimensions of the DCB configuration. We then illustrated the practical application of our theory using this law through three examples: Poly-Styrene foam, Poly-Urethane foam, and regular honeycomb. To do this, we utilised expressions and data on the micropolar behavior of these materials from existing literature, leading us to crucial findings concerning the theory's limits of applicability. Our scaling law brought to light a significant interplay of lengths between various dimensions of the DCB configuration and the micropolar characteristic length. This interplay dictates the behaviour of the configuration and stands as a decisive factor in the design process, guiding the selection of the optimal modelling theory.

\section*{Declaration of Competing Interest}
The authors declare that they have no known competing financial interests or personal relationships that could have appeared to influence the work reported in this paper.

\section*{CRediT authorship contribution statement}
A. E. F. Athanasiadis, M. K. Budzik, D. Fernando, and M. A. Dias performed the research and wrote the paper. A. E. F. Athanasiadis and M. A. Dias analysed the methodologies and results. M. A. Dias designed the research and acquired funds.

\section*{Acknowledgements} 
M. A. Dias would like to thank UKRI for support under the EPSRC Open Fellowship scheme (Project No. EP/W019450/1). M. K. Budzik acknowledges the Velux Foundations for support under the Villum Experiment program (VIL50302).

\printcredits

\bibliographystyle{model1-num-names}


\begin{thebibliography}{42}
\expandafter\ifx\csname natexlab\endcsname\relax\def\natexlab#1{#1}\fi
\providecommand{\url}[1]{\texttt{#1}}
\providecommand{\href}[2]{#2}
\providecommand{\path}[1]{#1}
\providecommand{\DOIprefix}{doi:}
\providecommand{\ArXivprefix}{arXiv:}
\providecommand{\URLprefix}{URL: }
\providecommand{\Pubmedprefix}{pmid:}
\providecommand{\doi}[1]{\href{http://dx.doi.org/#1}{\path{#1}}}
\providecommand{\Pubmed}[1]{\href{pmid:#1}{\path{#1}}}
\providecommand{\bibinfo}[2]{#2}
\ifx\xfnm\relax \def\xfnm[#1]{\unskip,\space#1}\fi
\bibitem[{Ashby(2016)}]{ashby2016}
\bibinfo{author}{M.~Ashby}, \bibinfo{title}{Materials Selection in Mechanical
  Design}, \bibinfo{publisher}{Elsevier Science}, \bibinfo{year}{2016}.
\bibitem[{Zheng et~al.(2014)Zheng, Lee, Weisgraber, Shusteff, DeOtte, Duoss,
  Kuntz, Biener, Ge, Jackson et~al.}]{zheng2014ultralight}
\bibinfo{author}{X.~Zheng}, \bibinfo{author}{H.~Lee}, \bibinfo{author}{T.~H.
  Weisgraber}, \bibinfo{author}{M.~Shusteff}, \bibinfo{author}{J.~DeOtte},
  \bibinfo{author}{E.~B. Duoss}, \bibinfo{author}{J.~D. Kuntz},
  \bibinfo{author}{M.~M. Biener}, \bibinfo{author}{Q.~Ge},
  \bibinfo{author}{J.~A. Jackson}, et~al.,
\newblock \bibinfo{title}{Ultralight, ultrastiff mechanical metamaterials},
\newblock \bibinfo{journal}{Science} \bibinfo{volume}{344}
  (\bibinfo{year}{2014}) \bibinfo{pages}{1373--1377}.
\bibitem[{Gibson and Ashby(1997)}]{gibson1997}
\bibinfo{author}{L.~J. Gibson}, \bibinfo{author}{M.~F. Ashby},
  \bibinfo{title}{Cellular Solids: Structure and Properties}, Cambridge Solid
  State Science Series, \bibinfo{edition}{2} ed., \bibinfo{publisher}{Cambridge
  University Press}, \bibinfo{year}{1997}.
  \DOIprefix\doi{10.1017/CBO9781139878326}.
\bibitem[{Deshpande et~al.(2001)Deshpande, Fleck, and Ashby}]{deshpande2001}
\bibinfo{author}{V.~Deshpande}, \bibinfo{author}{N.~Fleck},
  \bibinfo{author}{M.~Ashby},
\newblock \bibinfo{title}{Effective properties of the octet-truss lattice
  material},
\newblock \bibinfo{journal}{Journal of the Mechanics and Physics of Solids}
  \bibinfo{volume}{49} (\bibinfo{year}{2001}) \bibinfo{pages}{1747--1769}.
\bibitem[{Shaikeea et~al.(2022)Shaikeea, Cui, O'Masta, Zheng, and
  Deshpande}]{shaikeea2022}
\bibinfo{author}{A.~J.~D. Shaikeea}, \bibinfo{author}{H.~Cui},
  \bibinfo{author}{M.~O'Masta}, \bibinfo{author}{X.~R. Zheng},
  \bibinfo{author}{V.~S. Deshpande},
\newblock \bibinfo{title}{The toughness of mechanical metamaterials},
\newblock \bibinfo{journal}{Nature Materials} \bibinfo{volume}{21}
  (\bibinfo{year}{2022}) \bibinfo{pages}{297--304}.
\bibitem[{Cosserat and Cosserat(1909)}]{cosserat1909}
\bibinfo{author}{F.~Cosserat}, \bibinfo{author}{E.~Cosserat},
  \bibinfo{title}{Th{\'e}orie des corps d{\'e}formables, par E. Cosserat, ...
  F. Cosserat, ...}, \bibinfo{publisher}{A. Hermann et fils},
  \bibinfo{year}{1909}.
\bibitem[{Eringen(1968)}]{eringen1968}
\bibinfo{author}{A.~C. Eringen}, \bibinfo{title}{Theory of micropolar
  elasticity. In “Fracture, An Advanced Treatise” (H. Liebowitz, ed.), Vol.
  2, chapter 7}, \bibinfo{publisher}{Academic Press, New York},
  \bibinfo{year}{1968}.
\bibitem[{Nowacki(1986)}]{nowacki1986}
\bibinfo{author}{W.~Nowacki}, \bibinfo{title}{Theory of Asymmetric Elasticity},
  \bibinfo{publisher}{Pergamon Press}, \bibinfo{year}{1986}.
\bibitem[{Mindlin(1965)}]{mindlin1965}
\bibinfo{author}{R.~Mindlin},
\newblock \bibinfo{title}{Stress functions for a cosserat continuum},
\newblock \bibinfo{journal}{International Journal of Solids and Structures}
  \bibinfo{volume}{1} (\bibinfo{year}{1965}) \bibinfo{pages}{265--271}.
\bibitem[{Lakes(1983)}]{lakes1983}
\bibinfo{author}{R.~Lakes},
\newblock \bibinfo{title}{Size effects and micromechanics of a porous solid},
\newblock \bibinfo{journal}{Journal of Materials Science} \bibinfo{volume}{18}
  (\bibinfo{year}{1983}) \bibinfo{pages}{2572--2580}.
\bibitem[{Lakes(1986)}]{lakes1986}
\bibinfo{author}{R.~Lakes},
\newblock \bibinfo{title}{Experimental microelasticity of two porous solids},
\newblock \bibinfo{journal}{International Journal of Solids and Structures}
  \bibinfo{volume}{22} (\bibinfo{year}{1986}) \bibinfo{pages}{55--63}.
\bibitem[{Lakes(1991)}]{lakes1991}
\bibinfo{author}{R.~Lakes},
\newblock \bibinfo{title}{{Experimental Micro Mechanics Methods for
  Conventional and Negative Poisson’s Ratio Cellular Solids as Cosserat
  Continua}},
\newblock \bibinfo{journal}{Journal of Engineering Materials and Technology}
  \bibinfo{volume}{113} (\bibinfo{year}{1991}) \bibinfo{pages}{148--155}.
\bibitem[{Gauthier and Jahsman(1975)}]{gauthier1975}
\bibinfo{author}{R.~D. Gauthier}, \bibinfo{author}{W.~E. Jahsman},
\newblock \bibinfo{title}{{A Quest for Micropolar Elastic Constants}},
\newblock \bibinfo{journal}{Journal of Applied Mechanics} \bibinfo{volume}{42}
  (\bibinfo{year}{1975}) \bibinfo{pages}{369--374}.
\bibitem[{Adhikary and Dyskin(1997)}]{adhikary1997}
\bibinfo{author}{D.~Adhikary}, \bibinfo{author}{A.~Dyskin},
\newblock \bibinfo{title}{A cosserat continuum model for layered materials},
\newblock \bibinfo{journal}{Computers and Geotechnics} \bibinfo{volume}{20}
  (\bibinfo{year}{1997}) \bibinfo{pages}{15--45}.
\bibitem[{Bažant and Christensen(1972)}]{bazant1972}
\bibinfo{author}{Z.~Bažant}, \bibinfo{author}{M.~Christensen},
\newblock \bibinfo{title}{Analogy between micropolar continuum and grid
  frameworks under initial stress},
\newblock \bibinfo{journal}{International Journal of Solids and Structures}
  \bibinfo{volume}{8} (\bibinfo{year}{1972}) \bibinfo{pages}{327--346}.
\bibitem[{Chen et~al.(1998)Chen, Huang, and Ortiz}]{chen1998}
\bibinfo{author}{J.~Chen}, \bibinfo{author}{Y.~Huang},
  \bibinfo{author}{M.~Ortiz},
\newblock \bibinfo{title}{Fracture analysis of cellular materials: A strain
  gradient model},
\newblock \bibinfo{journal}{Journal of the Mechanics and Physics of Solids}
  \bibinfo{volume}{46} (\bibinfo{year}{1998}) \bibinfo{pages}{789--828}.
\bibitem[{Wang and Stronge(1999)}]{wang1998}
\bibinfo{author}{X.~L. Wang}, \bibinfo{author}{W.~J. Stronge},
\newblock \bibinfo{title}{Micropolar theory for two-dimensional stresses in
  elastic honeycomb},
\newblock \bibinfo{journal}{Proceedings: Mathematical, Physical and Engineering
  Sciences} \bibinfo{volume}{455} (\bibinfo{year}{1999})
  \bibinfo{pages}{2091--2116}.
\bibitem[{Kumar and McDowell(2004)}]{kumar2004}
\bibinfo{author}{R.~S. Kumar}, \bibinfo{author}{D.~L. McDowell},
\newblock \bibinfo{title}{Generalized continuum modeling of 2-d periodic
  cellular solids},
\newblock \bibinfo{journal}{International Journal of Solids and Structures}
  \bibinfo{volume}{41} (\bibinfo{year}{2004}) \bibinfo{pages}{7399--7422}.
\bibitem[{Berkache et~al.(2022)Berkache, Phani, and Ganghoffer}]{berkache2022}
\bibinfo{author}{K.~Berkache}, \bibinfo{author}{S.~Phani},
  \bibinfo{author}{J.-F. Ganghoffer},
\newblock \bibinfo{title}{Micropolar effects on the effective elastic
  properties and elastic fracture toughness of planar lattices},
\newblock \bibinfo{journal}{European Journal of Mechanics - A/Solids}
  \bibinfo{volume}{93} (\bibinfo{year}{2022}) \bibinfo{pages}{104489}.
\bibitem[{{Dos Reis} and Ganghoffer(2012)}]{dosreis2012}
\bibinfo{author}{F.~{Dos Reis}}, \bibinfo{author}{J.~Ganghoffer},
\newblock \bibinfo{title}{Construction of micropolar continua from the
  asymptotic homogenization of beam lattices},
\newblock \bibinfo{journal}{Computers \& Structures} \bibinfo{volume}{112-113}
  (\bibinfo{year}{2012}) \bibinfo{pages}{354--363}.
\bibitem[{Alavi et~al.(2022)Alavi, Ganghoffer, Sadighi, Nasimsobhan, and
  Akbarzadeh}]{alavi2022}
\bibinfo{author}{S.~Alavi}, \bibinfo{author}{J.~Ganghoffer},
  \bibinfo{author}{M.~Sadighi}, \bibinfo{author}{M.~Nasimsobhan},
  \bibinfo{author}{A.~Akbarzadeh},
\newblock \bibinfo{title}{Continualization method of lattice materials and
  analysis of size effects revisited based on cosserat models},
\newblock \bibinfo{journal}{International Journal of Solids and Structures}
  \bibinfo{volume}{254-255} (\bibinfo{year}{2022}) \bibinfo{pages}{111894}.
\bibitem[{Alavi et~al.(2020)Alavi, Sadighi, Pazhooh, and
  Ganghoffer}]{alavi2020}
\bibinfo{author}{S.~E. Alavi}, \bibinfo{author}{M.~Sadighi},
  \bibinfo{author}{M.~D. Pazhooh}, \bibinfo{author}{J.-F. Ganghoffer},
\newblock \bibinfo{title}{Development of size-dependent consistent couple
  stress theory of timoshenko beams},
\newblock \bibinfo{journal}{Applied Mathematical Modelling}
  \bibinfo{volume}{79} (\bibinfo{year}{2020}) \bibinfo{pages}{685--712}.
\bibitem[{Ha et~al.(2016)Ha, Plesha, and Lakes}]{ha2016}
\bibinfo{author}{C.~S. Ha}, \bibinfo{author}{M.~E. Plesha},
  \bibinfo{author}{R.~S. Lakes},
\newblock \bibinfo{title}{Chiral three-dimensional isotropic lattices with
  negative poisson's ratio},
\newblock \bibinfo{journal}{physica status solidi (b)} \bibinfo{volume}{253}
  (\bibinfo{year}{2016}) \bibinfo{pages}{1243--1251}.
\bibitem[{Rueger and Lakes(2018)}]{rueger2018}
\bibinfo{author}{Z.~Rueger}, \bibinfo{author}{R.~S. Lakes},
\newblock \bibinfo{title}{Strong cosserat elasticity in a transversely
  isotropic polymer lattice},
\newblock \bibinfo{journal}{Phys. Rev. Lett.} \bibinfo{volume}{120}
  (\bibinfo{year}{2018}) \bibinfo{pages}{065501}.
\bibitem[{Winkler(1868)}]{winkler}
\bibinfo{author}{E.~Winkler}, \bibinfo{title}{Die Lehre von der Elastizit{\"a}t
  und Festigkeit mit besonderer R{\"u}cksicht auf ihre Anwendung in der
  Technik: f{\"u}r polytechnische Schulen, Bauakademien, Ingenieure,
  Maschinenbauer, Architecten, etc}, \bibinfo{number}{v. 1},
  \bibinfo{publisher}{Dominicius}, \bibinfo{year}{1868}.
\bibitem[{Dillard et~al.(2018)Dillard, Mukherjee, Karnal, Batra, and
  Frechette}]{dillard2018}
\bibinfo{author}{D.~A. Dillard}, \bibinfo{author}{B.~Mukherjee},
  \bibinfo{author}{P.~Karnal}, \bibinfo{author}{R.~C. Batra},
  \bibinfo{author}{J.~Frechette},
\newblock \bibinfo{title}{A review of winkler{'}s foundation and its profound
  influence on adhesion and soft matter applications},
\newblock \bibinfo{journal}{Soft Matter} \bibinfo{volume}{14}
  (\bibinfo{year}{2018}) \bibinfo{pages}{3669--3683}.
\bibitem[{Allen(1969)}]{allen1969}
\bibinfo{author}{H.~G. Allen},
\newblock \bibinfo{title}{Chapter 8 - wringling and other forms of local
  instability},
\newblock in: \bibinfo{editor}{H.~G. Allen} (Ed.), \bibinfo{booktitle}{Analysis
  and Design of Structural Sandwich Panels}, The Commonwealth and International
  Library: Structures and Solid Body Mechanics Division,
  \bibinfo{publisher}{Pergamon}, \bibinfo{year}{1969}, pp.
  \bibinfo{pages}{156--189}. \URLprefix
  \url{https://www.sciencedirect.com/science/article/pii/B9780080128702500122}.
  \DOIprefix\doi{https://doi.org/10.1016/B978-0-08-012870-2.50012-2}.
\bibitem[{Audoly and Boudaoud(2008)}]{audoly2008a}
\bibinfo{author}{B.~Audoly}, \bibinfo{author}{A.~Boudaoud},
\newblock \bibinfo{title}{Buckling of a stiff film bound to a compliant
  substrate—part i:: Formulation, linear stability of cylindrical patterns,
  secondary bifurcations},
\newblock \bibinfo{journal}{Journal of the Mechanics and Physics of Solids}
  \bibinfo{volume}{56} (\bibinfo{year}{2008}) \bibinfo{pages}{2401--2421}.
\bibitem[{Soutis(2005)}]{soutis2005}
\bibinfo{author}{C.~Soutis},
\newblock \bibinfo{title}{Carbon fiber reinforced plastics in aircraft
  construction},
\newblock \bibinfo{journal}{Materials Science and Engineering: A}
  \bibinfo{volume}{412} (\bibinfo{year}{2005}) \bibinfo{pages}{171--176}.
  \bibinfo{note}{International Conference on Recent Advances in Composite
  Materials}.
\bibitem[{Thomas and Ramachandra(2018)}]{thomas2018}
\bibinfo{author}{L.~Thomas}, \bibinfo{author}{M.~Ramachandra},
\newblock \bibinfo{title}{Advanced materials for wind turbine blade- a review},
\newblock \bibinfo{journal}{Materials Today: Proceedings} \bibinfo{volume}{5}
  (\bibinfo{year}{2018}) \bibinfo{pages}{2635--2640}.
  \bibinfo{note}{International Conference on Advanced Materials and
  Applications (ICAMA 2016), June 15-17, 2016, Bengaluru, Karanataka, INDIA}.
\bibitem[{Alam et~al.(2018)Alam, Robert, and {Ó Brádaigh}}]{alam2018}
\bibinfo{author}{P.~Alam}, \bibinfo{author}{C.~Robert}, \bibinfo{author}{C.~M.
  {Ó Brádaigh}},
\newblock \bibinfo{title}{Tidal turbine blade composites - a review on the
  effects of hygrothermal aging on the properties of cfrp},
\newblock \bibinfo{journal}{Composites Part B: Engineering}
  \bibinfo{volume}{149} (\bibinfo{year}{2018}) \bibinfo{pages}{248--259}.
\bibitem[{Athanasiadis et~al.(2021)Athanasiadis, Dias, and
  Budzik}]{Athanasiadis2021}
\bibinfo{author}{A.~E. Athanasiadis}, \bibinfo{author}{M.~A. Dias},
  \bibinfo{author}{M.~K. Budzik},
\newblock \bibinfo{title}{Can confined mechanical metamaterials replace
  adhesives?},
\newblock \bibinfo{journal}{Extreme Mechanics Letters} \bibinfo{volume}{48}
  (\bibinfo{year}{2021}) \bibinfo{pages}{101411}.
\bibitem[{Hedvard et~al.(2024)Hedvard, Dias, and Budzik}]{hedvard2024}
\bibinfo{author}{M.~L. Hedvard}, \bibinfo{author}{M.~A. Dias},
  \bibinfo{author}{M.~K. Budzik},
\newblock \bibinfo{title}{Toughening mechanisms and damage propagation in
  architected-interfaces},
\newblock \bibinfo{journal}{International Journal of Solids and Structures}
  \bibinfo{volume}{288} (\bibinfo{year}{2024}) \bibinfo{pages}{112600}.
\bibitem[{Kanninen(1973)}]{kanninen1973}
\bibinfo{author}{M.~F. Kanninen},
\newblock \bibinfo{title}{An augmented double cantilever beam model for
  studying crack propagation and arrest},
\newblock \bibinfo{journal}{International Journal of Fracture}
  \bibinfo{volume}{9} (\bibinfo{year}{1973}) \bibinfo{pages}{83--92}.
\bibitem[{Heide-Jørgensen et~al.(2020)Heide-Jørgensen, Budzik, and
  Turner}]{heide2020}
\bibinfo{author}{S.~Heide-Jørgensen}, \bibinfo{author}{M.~K. Budzik},
  \bibinfo{author}{K.~T. Turner},
\newblock \bibinfo{title}{Mechanics and fracture of structured pillar
  interfaces},
\newblock \bibinfo{journal}{Journal of the Mechanics and Physics of Solids}
  \bibinfo{volume}{137} (\bibinfo{year}{2020}) \bibinfo{pages}{103825}.
\bibitem[{Budzik et~al.(2013)Budzik, Jumel, {Ben Salem}, and
  Shanahan}]{budzik2013}
\bibinfo{author}{M.~Budzik}, \bibinfo{author}{J.~Jumel},
  \bibinfo{author}{N.~{Ben Salem}}, \bibinfo{author}{M.~Shanahan},
\newblock \bibinfo{title}{Instrumented end notched flexure – crack
  propagation and process zone monitoring part ii: Data reduction and
  experimental},
\newblock \bibinfo{journal}{International Journal of Solids and Structures}
  \bibinfo{volume}{50} (\bibinfo{year}{2013}) \bibinfo{pages}{310--319}.
\bibitem[{{Ben Salem} et~al.(2014){Ben Salem}, Jumel, Budzik, Shanahan, and
  Lavelle}]{bensalem2014}
\bibinfo{author}{N.~{Ben Salem}}, \bibinfo{author}{J.~Jumel},
  \bibinfo{author}{M.~Budzik}, \bibinfo{author}{M.~Shanahan},
  \bibinfo{author}{F.~Lavelle},
\newblock \bibinfo{title}{Analytical and experimental investigations of crack
  propagation in adhesively bonded joints with the mixed mode bending (mmb)
  test part i: Macroscopic analysis \& digital image correlation measurements},
\newblock \bibinfo{journal}{Theoretical and Applied Fracture Mechanics}
  \bibinfo{volume}{74} (\bibinfo{year}{2014}) \bibinfo{pages}{209--221}.
\bibitem[{Christensen(1987)}]{christensen1987}
\bibinfo{author}{R.~M. Christensen},
\newblock \bibinfo{title}{{Sufficient Symmetry Conditions for Isotropy of the
  Elastic Moduli Tensor}},
\newblock \bibinfo{journal}{Journal of Applied Mechanics} \bibinfo{volume}{54}
  (\bibinfo{year}{1987}) \bibinfo{pages}{772--777}.
\bibitem[{Nakamura and Lakes(1995)}]{nakamura1995}
\bibinfo{author}{S.~Nakamura}, \bibinfo{author}{R.~Lakes},
\newblock \bibinfo{title}{Finite element analysis of saint‐venant end effects
  in micropolar elastic solids},
\newblock \bibinfo{journal}{Engineering Computations} \bibinfo{volume}{12}
  (\bibinfo{year}{1995}) \bibinfo{pages}{571--587}.
\bibitem[{Hoff and Mautner(1945)}]{hoff1945}
\bibinfo{author}{N.~J. Hoff}, \bibinfo{author}{S.~E. Mautner},
\newblock \bibinfo{title}{The buckling of sandwich-type panels},
\newblock \bibinfo{journal}{Journal of the Aeronautical Sciences}
  \bibinfo{volume}{12} (\bibinfo{year}{1945}) \bibinfo{pages}{285--297}.
\bibitem[{Gough et~al.(1940)Gough, Elam, Tipper, and De~Bruyne}]{gough1940}
\bibinfo{author}{G.~S. Gough}, \bibinfo{author}{C.~F. Elam},
  \bibinfo{author}{G.~H. Tipper}, \bibinfo{author}{N.~A. De~Bruyne},
\newblock \bibinfo{title}{The stabilisation of a thin sheet by a continuous
  supporting medium},
\newblock \bibinfo{journal}{The Aeronautical Journal} \bibinfo{volume}{44}
  (\bibinfo{year}{1940}) \bibinfo{pages}{12–43}.
\bibitem[{Suh et~al.(2020)Suh, Sun, and O’Connor}]{suh2020}
\bibinfo{author}{H.~S. Suh}, \bibinfo{author}{W.~Sun}, \bibinfo{author}{D.~T.
  O’Connor},
\newblock \bibinfo{title}{A phase field model for cohesive fracture in
  micropolar continua},
\newblock \bibinfo{journal}{Computer Methods in Applied Mechanics and
  Engineering} \bibinfo{volume}{369} (\bibinfo{year}{2020})
  \bibinfo{pages}{113181}.

\end{thebibliography}


\end{document}